# The ALMA Survey of Gas Evolution of PROtoplanetary Disks (AGE-PRO). IV. Dust and Gas Disk Properties in the Upper Scorpius Star-forming Region

Carolina Agurto-Gangas[1], Laura M. Pérez[1], Anibal Sierra[1,2], James Miley[3,4,5], Ke Zhang[6], Ilaria Pascucci[7],
Paola Pinilla[2], Dingshan Deng[7], John Carpenter[8], Leon Trapman[6], Miguel Vioque[9], Giovanni P. Rosotti[10],
Nicolas Kurtovic[11,12], Lucas A. Cieza[4,13], Rossella Anania[10], Benoît Tabone[14], Kamber Schwarz[11],
Michiel R. Hogerheijde[15,16], Estephani E. TorresVillanueva[6], Dary A. Ruiz-Rodriguez[8,17], and
Camilo González-Ruilova[4,5,13]

[1] Departamento de Astronomía, Universidad de Chile, Camino el Observatorio 1515, Las Condes, Santiago, Chile; cagurto@das.uchile.cl
[2] Mullard Space Science Laboratory, University College London, Holmbury St Mary, Dorking, Surrey RH5 6NT, UK
[3] Departamento de Física, Universidad de Santiago de Chile, Av. Victor Jara 3659, Santiago, Chile
[4] Millennium Nucleus on Young Exoplanets and their Moons (YEMS), Chile
[5] Center for Interdisciplinary Research in Astrophysics and Space Exploration (CIRAS), Universidad de Santiago, Chile
[6] Department of Astronomy, University of Wisconsin-Madison, 475 N Charter Street, Madison, WI 53706, USA
[7] Lunar and Planetary Laboratory, The University of Arizona, Tucson, AZ 85721, USA
[8] Joint ALMA Observatory, Avenida Alonso de Córdova 3107, Vitacura, Santiago, Chile
[9] European Southern Observatory, Karl-Schwarzschild-Str. 2, 85748 Garching bei München, Germany
[10] Dipartimento di Fisica, Università degli Studi di Milano, Via Celoria 16, I-20133 Milano, Italy
[11] Max-Planck-Institut fur Astronomie (MPIA), Konigstuhl 17, 69117 Heidelberg, Germany
[12] Max Planck Institute for Extraterrestrial Physics, Giessenbachstrasse 1, D-85748 Garching, Germany
[13] Instituto de Estudios Astrofísicos, Universidad Diego Portales, Av. Ejercito 441, Santiago, Chile
[14] Institut d'Astrophysique Spatiale, Université Paris-Saclay, CNRS, Bâtiment 121, 91405, Orsay Cedex, France
[15] Leiden Observatory, Leiden University, PO Box 9513, 2300 RA Leiden, The Netherlands
[16] Anton Pannekoek Institute for Astronomy, University of Amsterdam, The Netherlands
[17] National Radio Astronomy Observatory, 520 Edgemont Road, Charlottesville, VA 22903, USA
Received 2024 September 13; revised 2025 March 7; accepted 2025 March 7; published 2025 July 31

## Abstract

The Atacama Large Millimeter/submillimeter Array (ALMA) large program AGE-PRO explores protoplanetary disk evolution by studying gas and dust across various ages. This work focuses on 10 evolved disks in Upper Scorpius, observed in dust continuum emission, CO and its isotopologues, and $N_2H^+$ with ALMA Bands 6 and 7. Disk radii, from the radial location enclosing 68% of the flux, are comparable to those in the younger Lupus region for both gas and dust tracers. However, solid masses are about an order of magnitude below those in Lupus and Ophiuchus, while the dust spectral index suggests some level of dust evolution. These empirical findings align with a combination of radial drift, dust trapping, and grain growth into larger bodies. A moderate correlation between CO and continuum fluxes suggests a link between gas and dust content, through the increased scatter compared to younger regions, possibly due to age variations, gas-to-dust ratio differences, or CO depletion. Additionally, the correlation between $C^{18}O$ and $N_2H^+$ fluxes observed in Lupus persists in Upper Scorpius, indicating a relatively stable CO gas abundance over the Class II stage of disk evolution. In conclusion, the AGE-PRO survey of Upper Scorpius disks reveals intriguing trends in disk evolution. The findings point toward potential gas evolution and the presence of dust traps in these older disks. Future high-resolution observations are needed to confirm these possibilities and further refine our understanding of disk evolution and planet formation in older environments.

*Unified Astronomy Thesaurus concepts:* Submillimeter astronomy (1647); Astrochemistry (75); Protoplanetary disks (1300)

## 1. Introduction

The main goal of the Atacama Large Millimeter/submillimeter Array (ALMA) large program "ALMA survey of Gas Evolution of PROtoplanetary disks" (AGE-PRO; K. Zhang et al. 2025) is to trace the evolution of gas disk mass and size using a well-defined sample of 30 disks over a broad range of ages. Upper Scorpius (hereafter Upper Sco) is an important star-forming region for studies of disk evolution as it provides a snapshot of disk properties at an intermediate to old age (5–10 Myr; E. J. de Geus et al. 1989; T. Preibisch et al. 2002; M. J. Pecaut et al. 2012; K. Sullivan & A. L. Kraus 2021). Additionally, Upper Sco is one of the nearest star-forming regions with a mean distance of 142 pc (K. L. Luhman & T. L. Esplin 2020; C. A. L. Bailer-Jones et al. 2021; M. Fang et al. 2023), making it ideal for sensitive, photometric, and spatially resolved studies of the properties of evolved disks.

Submillimeter observations are essential for studying the cold dust and gas of circumstellar disks (C. F. Manara et al. 2023). Previous studies show that disks in older regions, such as Upper Sco, are fainter (J. M. Carpenter et al. 2014; S. A. Barenfeld et al. 2016, 2017) and more compact (N. Hendler et al. 2020) than their counterparts in younger regions. In particular, S. A. Barenfeld et al. (2016) showed strong evidence for systematically lower dust masses in Upper Sco, in comparison to those measured in the younger Taurus star-forming region, for a sample of 75 disks around low-mass stars (0.14–1.66 $M_\odot$). Although millimeter dust continuum emission is routinely detected, there is a lower rate of molecular

Table 1
Stellar Properties for the AGE-PRO Upper Sco Sample

| ID | 2MASS | R.A. | Decl. | SpT | Dist (pc) | $T_{eff}$ (K) | $L_\star$ ($L_\odot$) | $M_\star$ ($M_\odot$) | $\log_{10}(\dot{M})$ ($M_\odot$ yr$^{-1}$) | $A_v$ | Age (Myr) |
|---|---|---|---|---|---|---|---|---|---|---|---|
| 1 | J16120668-3010270 | $16^h12^m06\overset{s}{.}664$ | $-30^d10^m27\overset{s}{.}617$ | M0 | 131.9 | 3700 | 0.25 | $0.51^{+0.69}_{-0.39}$ | $-9.40$ | 0.056 | $4.5^{+0.9}_{-2.3}$ |
| 2 | J16054540-2023088 | $16^h05^m45\overset{s}{.}379$ | $-20^d23^m09\overset{s}{.}330$ | M4.5 | 137.6 | 3020 | 0.07 | $0.13^{+0.14}_{-0.05}$ | $-9.36$ | 0.262 | $2.0^{+0.9}_{-1.1}$ |
| 3 | J16020757-2257467 | $16^h02^m07\overset{s}{.}556$ | $-22^d57^m47\overset{s}{.}424$ | M2 | 139.6 | 3490 | 0.15 | $0.37^{+0.54}_{-0.28}$ | $-10.98$ | 0.523 | $3.6^{+0.7}_{-1.8}$ |
| 4 | J16111742-1918285 | $16^h11^m17\overset{s}{.}406$ | $-19^d18^m29\overset{s}{.}231$ | M0.25 | 136.9 | 3735 | 0.35 | $0.50^{+0.65}_{-0.38}$ | ⋯ | 0.896 | $2.1^{+0.4}_{-1.1}$ |
| 5 | J16145026-2332397 | $16^h14^m50\overset{s}{.}249$ | $-23^d32^m40\overset{s}{.}238$ | M3 | 144.0 | 3360 | 0.11 | $0.29^{+0.41}_{-0.19}$ | ⋯ | 1.445 | $3.8^{+0.9}_{-1.9}$ |
| 6 | J16163345-2521505 | $16^h16^m33\overset{s}{.}429$ | $-25^d21^m51\overset{s}{.}163$ | M0 | 158.4 | 3700 | 0.18 | $0.52^{+0.67}_{-0.39}$ | $-10.91$ | 1.069 | $5.8^{+1.1}_{-3.0}$ |
| 7 | J16202863-2442087 | $16^h20^m28\overset{s}{.}622$ | $-24^d42^m09\overset{s}{.}174$ | M2 | 152.7 | 3490 | 0.23 | $0.34^{+0.25}_{-0.47}$ | ⋯ | 1.710 | $1.8^{+0.3}_{-0.9}$ |
| 8 | J16221532-2511349 | $16^h22^m15\overset{s}{.}324$ | $-25^d11^m35\overset{s}{.}672$ | M3 | 139.0 | 3360 | 0.14 | $0.29^{+0.42}_{-0.20}$ | ⋯ | 1.903 | $2.1^{+0.4}_{-0.8}$ |
| 9 | J16082324-1930009 | $16^h08^m23\overset{s}{.}247$ | $-19^d30^m00\overset{s}{.}980$ | M0 | 137.0 | 3880 | 0.24 | $0.56^{+1.01}_{-0.63}$ | $-9.14$ | 0.888 | $6.6^{+1.2}_{-3.7}$ |
| 10 | J16090075-1908526 | $16^h09^m00\overset{s}{.}739$ | $-19^d08^m53\overset{s}{.}284$ | M0 | 136.9 | 3630 | 0.35 | $0.53^{+0.54}_{-0.31}$ | $-8.81$ | 1.156 | $1.5^{+0.3}_{-0.9}$ |

**Note.** Spectral types from C. F. Manara et al. (2020), K. L. Luhman (2022), and C. F. Manara et al. (2023). USco 2 was classified as M2 and later on as M4.5 by C. F. Manara et al. (2023) by fitting the spectrum. Distance, effective temperature, stellar luminosity, and mass accretion rate from C. A. L. Bailer-Jones et al. (2021), C. F. Manara et al. (2023), and M. Fang et al. (2023). We adopt a Bayesian framework as adopted in I. Pascucci et al. (2016) to estimate the stellar mass ($M_\star$) and ages from the young stellar objects tracks in the H-R diagram (Figure 2; G. A. Feiden (2016) tracks are adopted for stars with $T_{eff} > 3900$ K, and I. Baraffe et al. (2015) tracks are used for cooler objects), and the uncertainties for $M_\star$ and ages represent their 68% confidence interval.

gas detections in Upper Sco Class II disks (J. M. Carpenter et al. 2025).

Retrieving gas disk masses is challenging; one approach is to trace the disk mass with CO and its less abundant isotopologues. However, several physical and chemical processes in the disk affect these species and need to be taken into account, for example, selective photodissociation, freeze-out, CO depletion through drift, and/or conversion of CO into $CO_2$ ice on the surface of grains (e.g., A. Miotello et al. 2016; K. I. Öberg & E. A. Bergin 2016; M. Ruaud & U. Gorti 2019; S. Krijt et al. 2020; M. Ruaud et al. 2022). The complexity of disk thermochemical models results in different CO-inferred gas disk masses, and various levels of CO-to-$H_2$ depletion are needed to reconcile the disk mass values with the HD-inferred masses and expected gas-to-dust mass ratios (e.g., E. A. Bergin et al. 2013; C. Favre et al. 2013; M. K. McClure et al. 2016; K. Zhang et al. 2020; D. Deng et al. 2023; I. Pascucci et al. 2023). Complementary to CO, $N_2H^+$ observations can help solve this problem, as this molecule acts as an ionization tracer in disk midplanes that, together with $C^{18}O$, can be used to evaluate CO abundances and infer the gas disk mass, as shown in L. Trapman et al. (2022).

To systematically trace the evolution of protoplanetary disk mass and size across a range of ages, AGE-PRO employs a standardized observational approach, tracing dust continuum emission in Bands 6 (226 GHz, 1.33 mm) and 7 (285 GHz, 1.05 mm), CO and its isotopologues in Band 6, and $N_2H^+$ in Band 7. Since observations of molecular emission from $N_2H^+$ can be used to infer CO abundances (e.g., L. Trapman et al. 2022), this new data set enables a detailed and in-depth study of disk masses and sizes—for both gas and dust—quite relevant for a sample of older disks.

Another important aspect is that traditional viscous evolution models (J. E. Pringle 1981), which predict smooth disk dispersal, struggle to explain several key observations in Upper Sco. The observed reduction in dust mass over time, as reported by J. M. Carpenter et al. (2025), presents a challenge to these models, with possible interpretations including, but not limited to, rapid disk dispersal, the presence of compact, unresolved disks, and deviations from standard gas-to-dust mass ratios. These discrepancies highlight the limitations of purely viscous models and suggest the need for alternative mechanisms to drive disk evolution. Magnetohydrodynamical (MHD) disk wind models have shown that angular momentum loss through MHD winds lead to faster disk dispersal, higher gas-to-dust ratios, and can explain the observed compact disks (e.g., T. K. Suzuki et al. 2016; B. Tabone et al. 2022; I. Pascucci et al. 2023). Therefore, Upper Sco serves as a crucial test case to evaluate the relative importance of viscous processes and MHD winds in disk evolution, ultimately leading to a more comprehensive understanding of disk dispersal and its connection to stellar properties and dust evolution.

In this paper, we present ALMA AGE-PRO observations of the Upper Sco sample, placing them in context with previous studies and pursuing a first-order analysis. Further analysis and modeling of this data set, as well as comparisons between different regions, are presented in subsequent AGE-PRO publications (R. Anania et al. 2025; N. T. Kurtovic et al. 2025; J. M. Miley et al. 2025; L. Trapman et al. 2025a, 2025b; M. Vioque et al. 2025). In Section 2, we introduce the selected targets and observations. In Section 3, we present the data reduction process, including calibration and imaging. In Section 4, we describe the analysis and results. Finally, in Sections 5 and 6, we discuss the results and summary, respectively.

## 2. Targets and Observations

The AGE-PRO Upper Sco sample consists of 10 disks, and to ease identification, each one is given an ID number from 1 to 10 and the prefix "USco." The selection criteria are described in Section 2.1, and the stellar properties, summarized in Table 1, are obtained following the procedures described in Section 2.2. Besides the Two Micron All Sky Survey (2MASS) designation presented in Table 1, other names that may be given to these systems are listed in Table 9 of Appendix B.

### 2.1. Sample Selection in Upper Sco

Disks were selected from previous ALMA surveys of the region (S. A. Barenfeld et al. 2016, 2017; J. M. Carpenter et al.

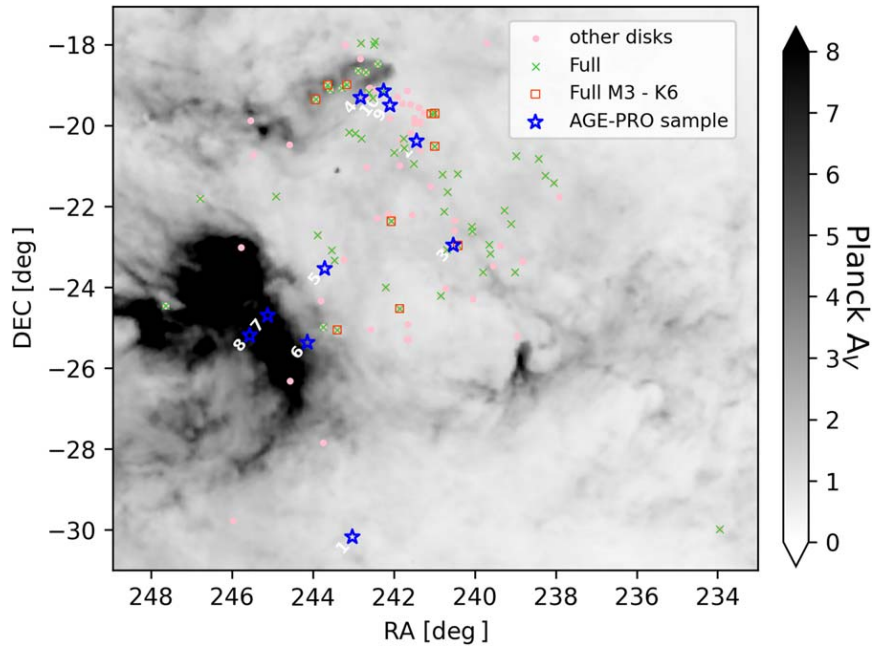

**Figure 1.** Class II disks in Upper Sco and the selected AGE-PRO sample on top of the Planck dust map (Planck Collaboration et al. 2014). The 10 Upper Sco disks selected in the AGE-PRO sample are shown as blue stars with their assigned ID in white, and the other disks are represented as pink points. On those pink points, green crosses show the Class II disks, and orange squares mark disks with a spectral type between M3 and K6.

2025), which covered ∼340 GHz continuum and $^{12}$CO ($J = 3$–2) lines in Band 7. Our selection criteria consider the following: (a) sources are of a stellar spectral type (SpT) between M3[M4.5]–K6, roughly corresponding to a stellar mass range of $M_\star \sim 0.3$–$0.8\,M_\odot$ according to theoretical models of evolutionary tracks of pre-main-sequence stars (e.g., I. Baraffe et al. 2015; G. A. Feiden 2016); (b) they are single or wide-separation binaries (>600 au); (c) they are confirmed members of Upper Sco, classified as "full disk" by K. L. Luhman & E. E. Mamajek (2012), K. L. Luhman & T. L. Esplin (2020), and K. L. Luhman (2022), selected to represent the disks that produce moderate to strong emission throughout the mid-infrared regime; (d) they are disks with previous detections of millimeter wavelength continuum emission and $^{12}$CO line emission (S. A. Barenfeld et al. 2016, 2017; J. M. Carpenter et al. 2025); and, finally, (e) known edge-on disks ($i > 70°$) are excluded.[18] Using these criteria, we selected 40 out of the 284 disk-bearing sources targeted with ALMA observations in J. M. Carpenter et al. (2025). The final sample was then chosen to cover the spread of disk continuum luminosities in the region 0.40–57.43 mJy at 0.89 mm (J. M. Carpenter et al. 2025). The only exception to the criteria described above is USco 2. Initially classified as M2 (±0.5 subclass) by K. L. Luhman et al. (2019) and S. A. Barenfeld et al. (2019), it was later classified spectroscopically as M4.5 (±0.5 subclass) by C. F. Manara et al. (2020). Since USco 2 satisfies all other criteria, and given the SpT uncertainty, we kept this source in the sample. Figure 1 shows the location of our targets in the Upper Sco star-forming region, with the Planck extinction map[19] (Planck Collaboration et al. 2014) in the background.

### 2.2. Stellar Parameters

Stellar masses for our Upper Sco sample were inferred by estimating the stellar effective temperature ($T_{\rm eff}$), to place each star in the Hertzsprung–Russell (H-R) diagram. Values for $T_{\rm eff}$ were inferred from the spectral type (C. F. Manara et al. 2020; K. L. Luhman 2022) using the $T_{\rm eff}$–SpT conversion in M. J. Pecaut & E. E. Mamajek (2013). We adopt an uncertainty of $\log T = 0.02$ (L. Hartmann 2001) in $T_{\rm eff}$, corresponding to a spectral type uncertainty of ±1 subclass. Visual extinction was estimated from the DENIS $I - J$ colors (DENIS Consortium 2005), the intrinsic colors in M. J. Pecaut & E. E. Mamajek (2013), and the J. A. Cardelli et al. (1989) extinction law as implemented by K. Barbary (2017) in Python. A minimum uncertainty for $A_v$ of 0.5 was used, following L. Hartmann (2001). The luminosities were then calculated from the dereddened 2MASS $J$-band magnitudes (R. M. Cutri et al. 2003; M. F. Skrutskie et al. 2006), the bolometric corrections in M. J. Pecaut & E. E. Mamajek (2013), and the Gaia-based distances in C. A. L. Bailer-Jones et al. (2021). Figure 2 shows the H-R diagram for our Upper Sco targets, with evolutionary tracks for pre-main-sequence stars from G. A. Feiden (2016) and I. Baraffe et al. (2015). To estimate stellar masses and ages from these evolutionary tracks, we follow the Bayesian inference approach developed by I. Pascucci et al. (2016). Our adopted procedure is available in the Python package ysoisochrone[20] (D. Deng et al. 2025a). In summary, for objects with $T_{\rm eff} \leqslant 3900$ K, we employ the evolutionary tracks of I. Baraffe et al. (2015), while for hotter stars, we transition to the tracks of G. A. Feiden (2016). The best-fit ages inferred for our Upper Sco sample are within 1–8 Myr. For a detailed discussion about the ages of the full AGE-PRO sample and their uncertainties, see K. Zhang et al. (2025). The stellar masses and ages constrained for each target are reported in

---
[18] Given the limited resolution of prior studies, disk inclinations were uncertain. Our new observations identify two systems exhibiting significantly inclined disks.
[19] Map created with Python package dustmaps by G. Green (2018).
[20] https://github.com/DingshanDeng/ysoisochrone

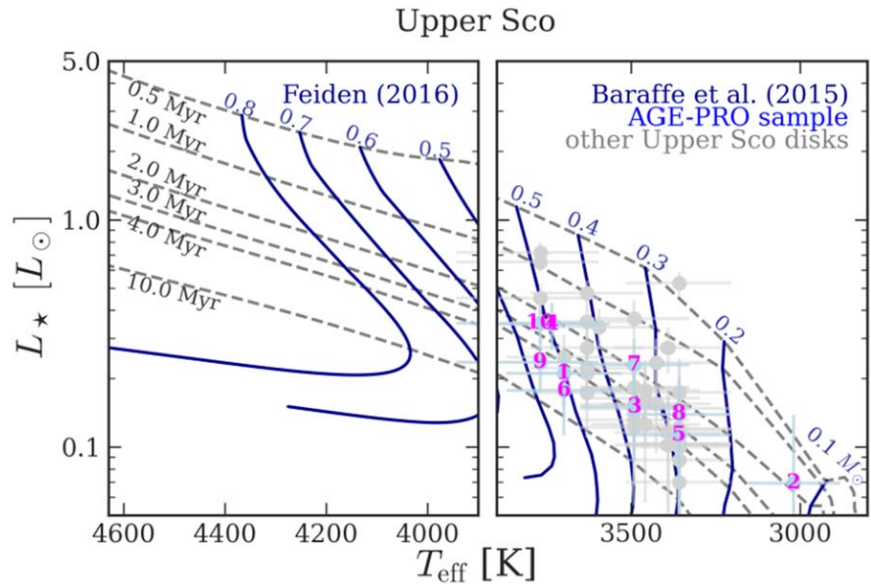

**Figure 2.** The Hertzsprung–Russell diagram for young stars in Upper Sco shown in gray, with AGE-PRO Upper Sco targets shown as blue points with magenta labels for their IDs. Their stellar luminosity and stellar effective temperature are summarized in Table 1. The evolutionary tracks for pre-main-sequence stars from G. A. Feiden (2016) and I. Baraffe et al. (2015) are presented in blue solid (for stellar masses from 0.1 to 0.8 $M_\odot$) and dark gray dashed lines (for ages from 0.5 to 10 Myr).

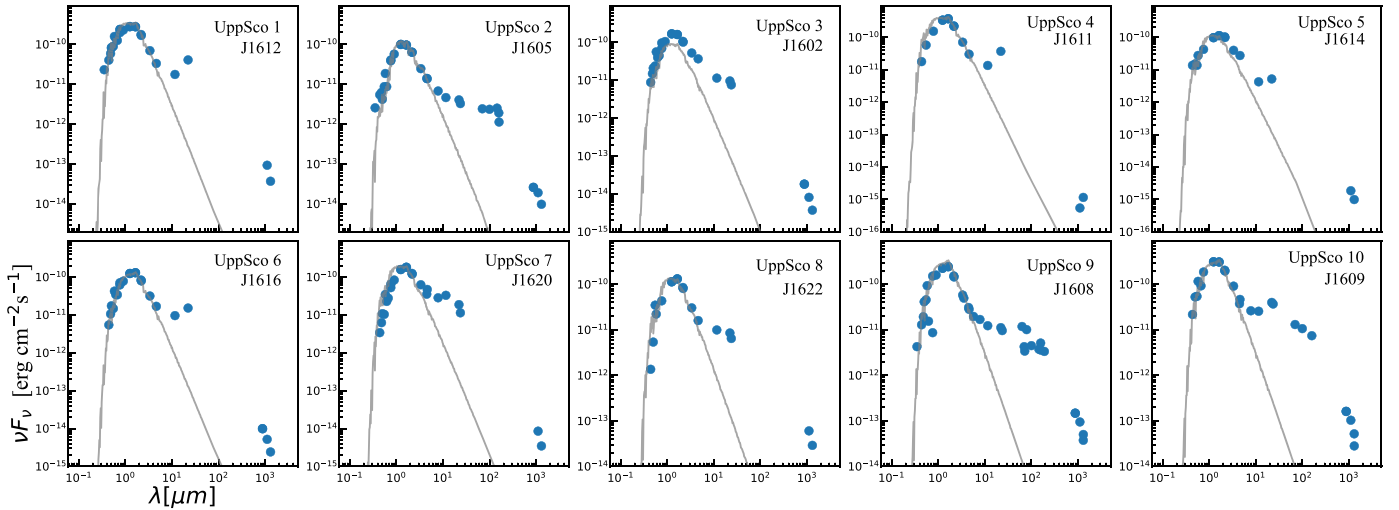

**Figure 3.** Spectral energy distributions for the AGE-PRO Upper Sco sample. The lines are the corresponding stellar photospheric model spectra from PHOENIX models (T. O. Husser et al. 2013). We apply the extinction to the photospheric models and then scale it to the distance of each star. The blue points are the observations collected from the literature: Sloan Digital Sky Survey (K. G. Stassun et al. 2019), APASS data release (DR) 9 (A. A. Henden et al. 2015), Gaia DR2/DR3 (Gaia Collaboration et al. 2018, 2021), 2MASS (R. M. Cutri et al. 2003; N. Zacharias et al. 2004a, 2004b), Spitzer (J. M. Carpenter et al. 2009; K. L. Luhman & E. E. Mamajek 2012; N. J. I. Evans et al. 2003), Wide-field Infrared Survey Explorer (B. M. Lasker et al. 2008), Herschel (G. S. Mathews et al. 2013), ALMA (S. A. Barenfeld et al. 2016; L. Testi et al. 2022), and Submillimeter Array (L. A. Cieza et al. 2008).

Table 1. The $M_\star$ estimated in this work are consistent with those in the literature (e.g., C. F. Manara et al. 2023).

Spectral energy distributions for all the disks in the Upper Sco sample are presented in Figure 3 and are compared with the Phoenix stellar photospheric models (T. O. Husser et al. 2013), where the extinction on each target has been applied accordingly, considering the low visual extinction in the Upper Sco region ($A_v < 2$ mag; J. M. Carpenter et al. 2014).

The distribution of two important properties—the stellar mass accretion rate (from M. Fang et al. 2023) and the dust continuum emission at 0.89 mm (from S. A. Barenfeld et al. 2016)—is presented in Figure 4. In this figure, the gray histograms correspond to Upper Sco targets classified as "full disks" from J. M. Carpenter et al. (2025), the orange histograms are for the subset of 40 "full disks" with SpT from M3 to K6, and the blue histograms include those in our AGE-PRO sample with known properties. Kolmogorov–Smirnov (K-S) tests are implemented to compare our AGE-PRO sample with the SpT-restricted sample. For both the stellar mass accretion rate and millimeter emission, we find no significant difference between our AGE-PRO sample and the "full" disk population surrounding similar ∼M3–K6 stars, as indicated by a $p$-value close to unity. This suggests that the AGE-PRO sample in Upper Sco exhibits a high degree of similarity to the overall stellar mass accretion and millimeter flux distribution in this region. As a test, we restrict the SpT range further, to M0–M4, which probes better the SpT of the AGE-PRO Upper Sco sample. We find that, in terms of accretion rate, our sample is representative ($p$-value close to 1), but since the dust continuum emission exhibits a dependence on spectral type, we find a much lower $p$-value, equal to 0.626, when comparing to the distribution of millimeter flux. This is consistent with the expectation that later spectral-type stars tend to exhibit a higher frequency of disks with lower millimeter fluxes.

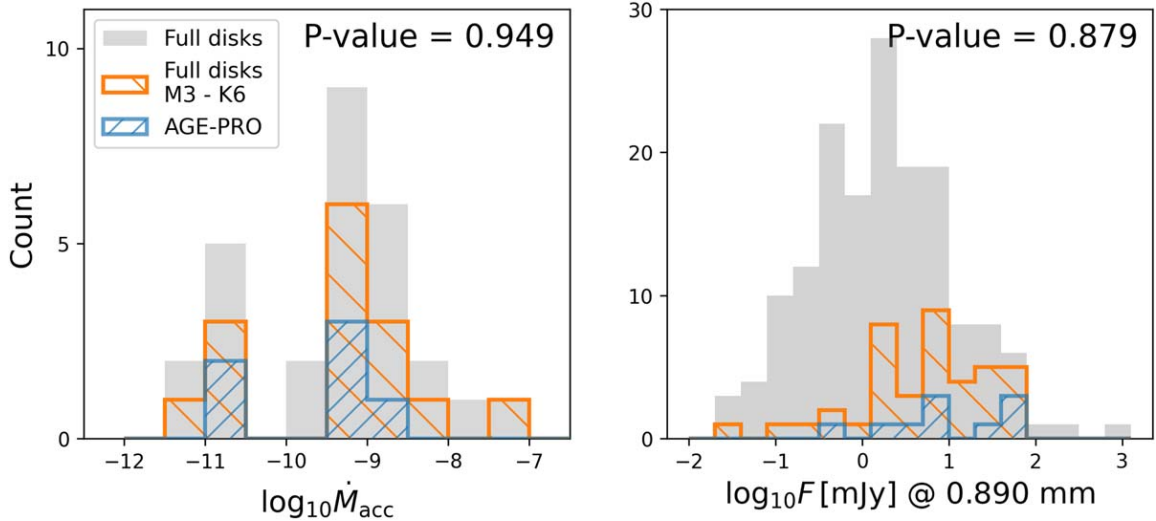

**Figure 4.** Histograms for the mass accretion rate ($\dot{M}_{\rm acc}$, from M. Fang et al. 2023, left) and millimeter flux ($F_{\rm 0.89mm}$, from S. A. Barenfeld et al. 2016; J. M. Carpenter et al. 2025, right) comparing our AGE-PRO Upper Sco sample (blue) with the "full disks" in Upper Sco that have SpT from M3 to K6 (orange), and all of the Upper Sco systems classified as "full disks" from J. M. Carpenter et al. (2025). A K-S test between the AGE-PRO and the SpT-restricted "full disk" sample finds $p$-values close to 1, suggesting they come from the same parent distribution.

### 2.3. Observations

The AGE-PRO Upper Sco observations were executed during 2022 March and 2022 July. In Band 6, the short baseline (SB) and long baseline (LB) observations were obtained in ALMA configurations C43-2 and C43-5, respectively, while only SB observations in configuration C43-3 were obtained in Band 7, for USco 1 through 8. Archival SB observations in Band 7 obtained in configuration C43-2 were employed in this study for USco 9 and USco 10 (see the observing log in Table 7, Appendix A).

For Band 6, the sample was divided into two clusters of five sources; the first subsample had two SB executions and five LB executions, while the other subsample had two SB executions and four LB executions. The combination of SB and LB observations allows us to measure the gas outer radius with a $3\sigma$ uncertainty down to 10–22 au at the Upper Sco distance, while facilitating a direct comparison with the observational setup for Lupus and Ophiuchus. Observations generally had $T_{\rm sys} \sim 69$–102.5 K, and rms phase variations after water vapor radiometer corrections were within the range of 6.1°–15.9°. In Band 6, we reach $\sim$50 minutes on-source observations and 30–50 minutes of calibration, with longer calibration times for LB observations. For the new Band 7 observations, sources were divided into two clusters, one with a single target (USco 1) and two SB executions, the other cluster with seven targets and 11 SB executions. On average, each target had 40–50 minutes on-source and $\sim$50 minutes of calibration. Additional details can be found in Table 7.

The correlator was set up with six and seven spectral windows (SPWs) for Bands 7 and 6, respectively. Two SPWs are for a continuum with 1.4 km s$^{-1}$ velocity resolution, and the other SPWs are designed to have a higher velocity resolution of 0.09–1.55 km s$^{-1}$ in Band 6 and 0.07–1.40 km s$^{-1}$ in Band 7. One of the highest resolution setups at 0.092 km s$^{-1}$ was used for the $^{12}$CO line, to enable kinematic studies. The observational setup employed in this study can be found in Table 2, and a list of the execution unique identifier's (UIDs) in the analysis is presented in Table 8 of Appendix A.

For USco targets 1, 5, 6, 7, and 8, one execution block in Band 6 (uid://a002/xf6d51d/x931e) was not included in the final imaging procedure as it was classified as "semipass" by ALMA quality assurance. Specifically, the flux calibrator (J1427-4206) used in this execution differed from those employed in the other two SB executions, and no flux density measurements of it were available near the time of our observations. Consequently, the SB execution associated with this particular flux calibrator is excluded from our analysis here.

In the case of USco 7, a millimeter-wave flare was detected during the time of SB observations. These SB executions while flaring were not included in the final continuum images and were not used for self-calibration of this target. However, we did include these executions in the molecular line imaging, since the line fluxes were consistent throughout the different epochs. All LB executions are included as normal. More details about this source are given in Section 4.6.2, and further analysis is presented in J. M. Miley et al. (2025).

## 3. Data Reduction

### 3.1. Calibration Strategy

The execution blocks were individually pipeline-calibrated using the CASA `v6.2.1-7 pipeline` for both SB and LB Band 6 and Band 7 data, and the `5.6.1-8 pipeline` for archival Band 7 data. Prior to self-calibration, we adhered to the standard procedures described in K. Zhang et al. (2025), which are astrometric and flux-scale alignment. For astrometric alignment, we align each of the SB and LB executions to a common phase center, listed for each source in Table 1. For flux-scale alignment, we check that different executions have a consistent flux within 10%. Only in the rare case of a flux-scale difference did we rescale the flux to the data set that was closer in time with an amplitude flux calibrator measurement from the ALMA calibrator monitoring.[21]

Table 2
Observational Setup in ALMA Bands 6 and 7

| Setup | Center Freq. (GHz) | Lines Targeted | Resolution (km s$^{-1}$) | Bandwidth (MHz) |
|---|---|---|---|---|
| B6 | 217.238530 | DCN J=3-2 | 1.552 | 58.59 |
|  | 218.000000 | Continuum | 1.552 | 1875 |
|  | 218.222192 | H$_2$CO (3$_{03}$–2$_{02}$) | 1.552 | 58.59 |
|  | 219.560358 | C$^{18}$O 2–1 | 0.096 | 58.59 |
|  | 220.398684 | $^{13}$CO 2–1 | 0.096 | 58.59 |
|  | 230.538000 | CO J = 2–1 | 0.092 | 58.59 |
|  | 231.321828 | N$_2$D$^+$ J = 3–2 | 0.091 | 58.59 |
|  | 234.019 | Continuum | 1.446 | 1875 |
| B7 | 278.200000 | Continuum | 1.216 | 1875 |
|  | 279.511760 | N$_2$H$^+$ $v = 0$ J = 3–2 | 0.076 | 1875 |
|  | 288.143858 | DCO$^+$ $v = 0$ J = 4–3 | 0.147 | 58.59 |
|  | 289.209066 | C$^{34}$S $v = 0$ J = 6–5 | 0.146 | 58.59 |
|  | 289.644907 | DCN $v = 0$ J = 4–3 | 0.146 | 58.59 |
|  | 290.850000 | Continuum, H$_2$CO (4$_{04}$–3$_{03}$) | 1.164 | 1875 |

For self-calibration, we also follow the standard procedures described in K. Zhang et al. (2025). In brief, we combined all SPWs using combine="spws" in the CASA gaincal task to improve the signal-to-noise ratio (SNR). For solution intervals longer than the scan length, we also combined scans using combine="spws,scans". Imaging with the tclean task was performed using a Briggs robust parameter of 0.5 and an elliptical mask around each source.

First, iterations of phase-only self-calibration were performed. We typically started with a large solution interval (e.g., solint='300s' or solint='inf'). If the peak intensity and image SNR improved by a few percent after an iteration of self-calibration, we proceeded with the next one, decreasing the solution time interval by a factor of ∼2.0. Then, amplitude self-calibration was attempted using a large solution interval close to the scan length, but solutions were applied only if the criteria outlined above were met. The self-calibration of the Band 6 data sets resulted in an average of ∼40% SNR improvement in most of the disks (no self-calibration was applied to USco 4 due to a low continuum SNR).

The same astrometric alignment, flux-scale offsets, and calibration tables inferred from the continuum visibilities were applied to the visibilities of molecular lines. With the task uvcontsub, the continuum emission was subtracted from the line data. Selecting the appropriate frequency ranges, we created continuum-subtracted data sets for each line and source. We imaged spectral lines with and without the above corrections and confirmed that the corrections improve their SNR. The data products and continuum and line measurement sets are available for download from the AGE-PRO repository.

### 3.2. Imaging Strategy

The main goals of imaging the Upper Sco data are to obtain measurements of (a) the dust continuum flux density and radial extent at both Bands 6 and 7, (b) the $^{12}$CO (2–1) flux and radial extent, and (c) the fluxes of the $^{13}$CO (2–1), C$^{18}$O (2–1), and N$_2$H$^+$ (3–2) molecular lines. As a secondary goal, part of the bandwidth was also used to target other lines, on which we also performed line imaging and flux measurements if detected: H$_2$CO (3$_{03}$–2$_{02}$), DCN (3–2), and N$_2$D$^+$ (3–2) in Band 6, and DCO$^+$ (4–3), C$^{34}$S (6–5), DCN (4–3), and H$_2$CO (4$_{04}$–3$_{03}$) in Band 7. Further details on our imaging procedure are provided in the AGE-PRO overview publication (K. Zhang et al. 2025).

#### 3.2.1. Continuum Imaging

The continuum images in Bands 6 and 7 are generated during self-calibration. For the final continuum image product, the tclean task with an elliptical mask, a Briggs robust parameter of 0.5, was used to achieve the desired angular resolution of ∼0.3″ for Band 6 and ∼0.7″ for Band 7.

#### 3.2.2. Line Imaging

Before line imaging, each measurement set was regridded onto a common velocity grid using the cvel2 task with a velocity spacing of 0.1 km s$^{-1}$ for $^{12}$CO and 0.2 km s$^{-1}$ for all other lines in Bands 6 and 7. To capture all the disk gas emission and overlap with the systemic velocity of the Upper Sco targets of $v_{sys}$ ∼ 2–7 km s$^{-1}$, we selected a wide range of channels, from −10.0 km s$^{-1}$ to +20 km s$^{-1}$.

Line image cubes were created following the procedure described in K. Zhang et al. (2025).[22] For imaging with the tclean task, we adopted a Keplerian rotation mask as the cleaning mask, constructed using the software from R. Teague (2020). The parameters adopted for these Keplerian masks are presented in Table 3 and include stellar mass, disk geometry, and the outer radius of the gas disk, which is selected to be large in order to incorporate all $^{12}$CO emission. Note that these values create Keplerian masks that encompass all observed emission and are similar to constrained values elsewhere. However, they are just reference values and are not meant to be taken as the actual geometry or stellar mass of the systems studied.

In order to achieve a spatial resolution similar to that of the continuum images, ∼0.25″–0.35″, while maintaining a sufficient SNR in the line image cubes, a dual approach to image reconstruction is adopted. For the continuum and $^{12}$CO, we used a robust parameter of 0.5. In the case of other lines and CO isotopologues, such as, for example, C$^{18}$O and N$_2$H$^+$, a robust value of 1.0 may be used to ensure detections of fainter lines at the expense of a ∼15% larger beam size. The line emission was cleaned down to 1 × rms, estimated in a line-free region. Finally, to constrain the dust and gas disk radii, the continuum images and molecular line cubes require a circularized beam. Circularization was first performed with uvtaper in the $uv$-plane to reach a beam circularized at more than 90%, and then we applied imsmooth to make a perfectly circularized beam.

---

[21] https://agepro.das.uchile.cl/

[22] All updated functions useful for data reduction and used in previous projects (S. M. Andrews et al. 2018; K. I. Öberg et al. 2021) are available on the project's website: https://agepro.das.uchile.cl

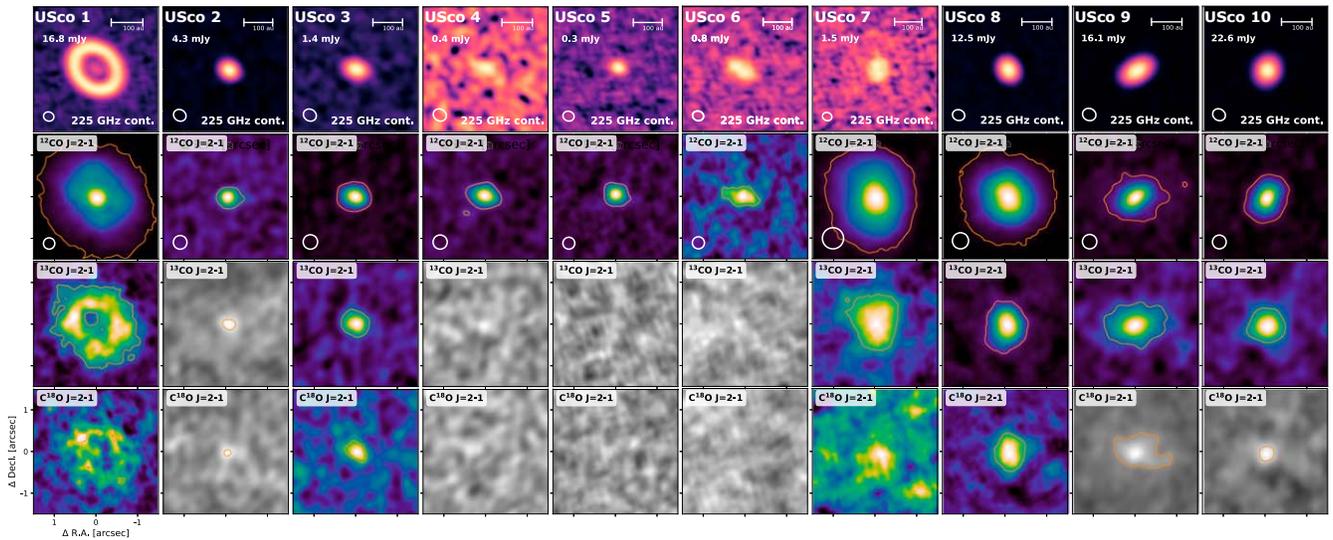

**Figure 5.** Gallery of Band 6 dust continuum images and line moment zero maps for our Upper Sco sample. The $^{12}$CO (2–1) images were obtained using robust 0.5 with a circularized beam, while the $^{13}$CO (2–1) and C$^{18}$O (2–1) images use robust 1.0 with a circularized beam. Contours in colored maps show $5\sigma$ detections, while contours in grayscale maps show either $3\sigma$ or nondetections. The continuum color scale is normalized to peak intensity.

## 4. Analysis and Results

### 4.1. Continuum Observations in Bands 6 and 7

Continuum images from Band 6 are presented in the top row of Figure 5. USco 1, the brightest disk in the sample, is the only source well resolved by these observations, showing a depleted gap in the image. In contrast, the rest of the sample is marginally resolved (USco 3, 6, 7, 8, 9, and 10) or unresolved (USco 2, 4, and 5). The typical rms value for the Band 6 continuum is 0.015 mJy beam$^{-1}$ and all the sources are detected with SNR ⩾ 9. Detailed analysis of the continuum visibilities, presented in M. Vioque et al. (2025), revealed some degree of substructures in six of the Upper Sco disks (USco 1, 6, 7, 8, 9, and 10), while USco 3 is resolved but without substructures at this resolution. The Band 7 continuum images are presented in the top row of Figure 8, revealing unresolved emission for most disks. Only USco 1 is resolved, with a subtle central depression of its emission, while in the field of view of USco 4, the emission of a background object, likely a galaxy, can be seen to the north. The typical rms value for the Band 7 continuum is 0.019 mJy beam$^{-1}$, and all the sources are detected with SNR ⩾ 11.

### 4.2. Spectral Line Observations in Band 6

In the second row of Figure 5, we present the moment zero maps of $^{12}$CO (2–1) integrated emission, while the third and

**Table 3**
Reference Disk Geometry

| Disk | Continuum[a] | | Keplerian Mask[b] | | | | |
|---|---|---|---|---|---|---|---|
| | incl (deg) | PA (deg) | incl (deg) | PA (deg) | $v_{\rm sys}$ (m s$^{-1}$) | $z_0$ (arcsec) | $M_\star$ ($M_\odot$) |
| USco 1 | 36.5 | 43.1 | 36 | 44 | 4526 | 0.17 | 0.73 |
| USco 2 | 51.0 | 63.0 | 51 | 63 | 3000 | 0.2 | 0.15 |
| USco 3 | 70.4 | 78.3 | 70 | 80 | 4400 | 0.3 | 0.85 |
| USco 4 | 80.6 | 70.9 | 80 | 260 | 4200 | 0.1 | 0.70 |
| USco 5 | 55.5 | 71.0 | 38 | 129 | 5551 | ⋯ | 0.40 |
| USco 6 | 62.0 | 62.0 | 62 | 242 | 6000 | 0.2 | 0.51 |
| USco 7 | 27.0 | 71.7 | 44 | 182 | 3931 | ⋯ | 0.50 |
| USco 8 | 52.0 | 20.0 | 52 | 25 | 3400 | 0.2 | 0.50 |
| USco 9 | 71.0 | 123.0 | 71 | 303 | 5100 | 0.2 | 0.65 |
| USco 10 | 50.3 | 155.3 | 50 | 327 | 4200 | 0.2 | 0.60 |

**Notes.** Values listed here are for reference only, to create a Keplerian mask for spectral line imaging for example, and should not be taken as the actual stellar parameters or disk geometry.
[a] Continuum geometry comes from fitting a 2D Gaussian to the Band 6 continuum image of each disk. The first two columns are the fitting inclination (incl) and position angle (PA), respectively.
[b] Parameters used for creating a Keplerian mask used in the spectral line imaging of each disk. The value of $z_0$ corresponds to the emission height at a radius of 1″.

bottom rows present integrated emission maps for $^{13}$CO (2–1) and C$^{18}$O (2–1), respectively, for our entire Upper Sco sample. All Upper Sco sources have $^{12}$CO (2–1) detections, which verifies our selection criteria. Emission from $^{13}$CO (2–1) is detected in USco 1, 3, 7, 8, 9, and 10, and is more compact than its $^{12}$CO (2–1) integrated emission. In the case of USco 2 and 6, which are some of the faintest disks in our AGE-PRO sample, $^{13}$CO is detected with an SNR higher than $3\sigma$ but lower than $5\sigma$. The C$^{18}$O emission exhibits an SNR $> 5\sigma$ in USco 1, 3, 7, and 8. For USco 2, 9, and 10, a centrally peaked $^{13}$CO emission is observed, but is only marginally detected ($3\sigma$). Neither $^{13}$CO or C$^{18}$O have been detected in USco targets 4, 5, and 6. Additionally, a detection of H$_2$CO ($3_{03}$–$2_{02}$) is clear for USco 1, 3, 8, and 9, while DCN (3–2) is only detected in USco 10.

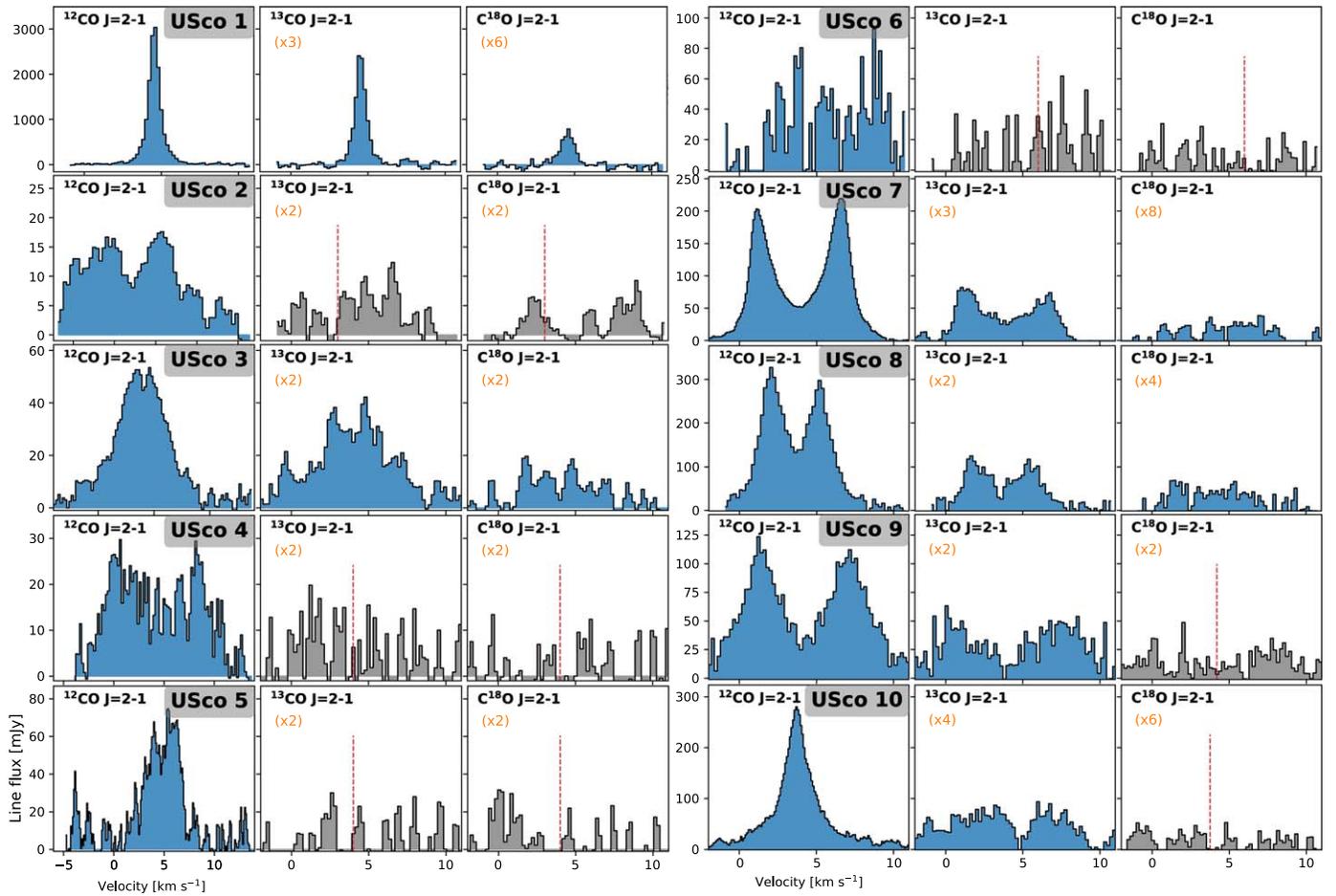

**Figure 6.** Disk-integrated line spectra of CO and its isotopologues for Upper Sco targets in Band 6. We use an aperture corresponding to the gas disk radii for $^{12}$CO (2–1), and smaller apertures for $^{13}$CO (2–1) and C$^{18}$O (2–1). Disks without >3$\sigma$ detections are shown in grayscale the red dashed line shows their systemic velocity identified from the $^{12}$CO (2–1) line.

The moment zero maps for these weaker lines, including nondetections, are presented in Figure E1 in Appendix E.

We generate velocity-stacked spectra, presented in Figure 6, where the line emission is deprojected according to the Keplerian velocity field using the method by H.-W. Yen et al. (2016). Even the velocity-stacked spectra fail to yield robust detections for those systems with 3$\sigma$ < SNR < 5$\sigma$; these cases are treated as upper limits.

We generate continuum and CO isotopologue radial profiles, presented in Figure 7, deprojecting by the disk geometry using GoFish[23] (R. Teague 2019). From those radial profiles, we obtain $R_{68}$ and $R_{90}$, the radial location where the flux is 68% and 90% of the total flux measured at the curve-of-growth (COG) peak (the COG method is described in Section 4.4). We note that Figures 6 and 7 were obtained by deprojecting using the disk geometry (inclination angle and position angle summarized in Table 3), which boosts the SNR of the line emission.

### 4.3. Spectral Line Observations in Band 7

The AGE-PRO Band 7 observations focus on the N$_2$H$^+$ (3–2) line and also include other weak lines: DCO$^+$ (4–3), C$^{34}$S (6–5), DCN (4–3), and H$_2$CO (4$_{04}$–3$_{03}$). Line emission from N$_2$H$^+$ (3–2) is detected in four disks (USco 1, 7, 9, and 10) with SNR > 5$\sigma$. For the rest of the sample, this line is not detected. Line emission from H$_2$CO (4$_{04}$–3$_{03}$) is detected in USco 1, 2, 3, 7, and 8. Additionally, USco 7 and 8 have DCN (4–3) detections, while USco 8 is the only source with a 5$\sigma$ detection in DCO$^+$ (4–3). Finally, line emission from C$^{34}$S (6–5) is partially detected in USco 1, with an SNR of 3$\sigma$. Figure 8 shows our Band 7 continuum images for all Upper Sco disks and circularized moment zero maps for all targeted molecules, while the velocity-stacked spectra are presented in Figure 9.

### 4.4. Constraints in Total Flux and Disk Size

We measure the total flux and disk radius for both continuum and molecular line emission using the COG method. This approach follows established methods in the literature (e.g., M. Ansdell et al. 2016, 2018; S. M. Andrews et al. 2018; E. Sanchis et al. 2021); further details of its application to AGE-PRO images are in Section 4.1 of K. Zhang et al. (2025), and here we describe it briefly. First, moment zero maps of each molecular line are made with bettermoments[24] using a Keplerian mask that contains all disk emission, (1.5 × $R_{\rm max}$), where $R_{\rm max}$ is defined as the largest distance where emission is detected over 1$\sigma$ in the individual channels. Then, we create elliptical apertures considering the disk inclination and position

---
[23] https://github.com/richteague/gofish/tree/master
[24] https://github.com/richteague/bettermoments

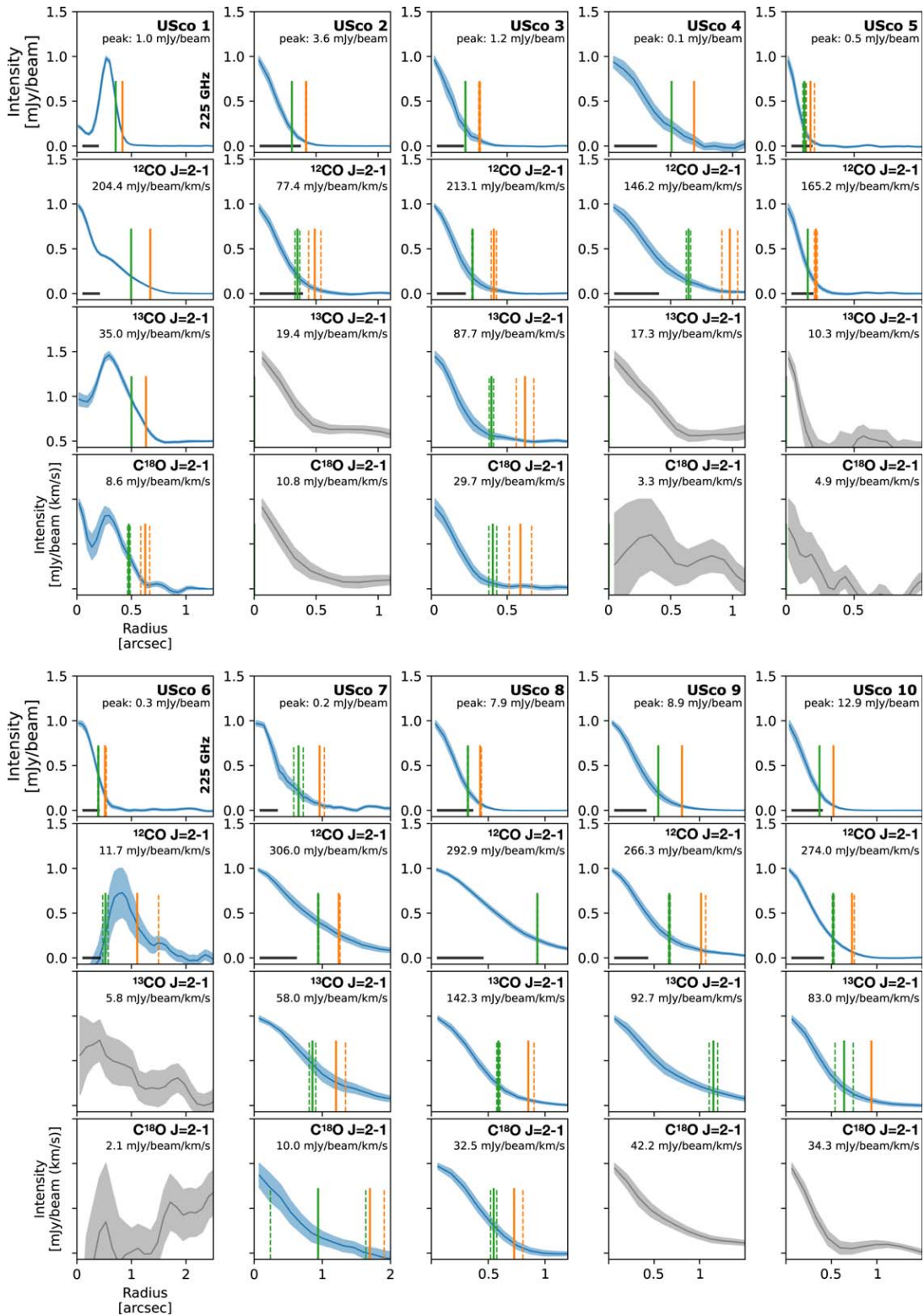

**Figure 7.** Normalized radial profiles of dust continuum and gas line emission for our Upper Sco sample in AGE-PRO. Profiles were obtained using GoFish; the deprojection employed the stellar parameters in Table 1 and disk geometries in Table 3. Orange and green continuous vertical lines mark the $R_{68\%}$ and $R_{90\%}$ radii, respectively; their uncertainties are shown by dashed lines. The beam size is presented in the continuum and $^{12}$CO panels as a black horizontal line in the lower-left corner of each panel. Radial profiles of nondetections are shown in gray.

angles. The flux inside each aperture, in both continuum and line emission, is measured to compile the COG as a function of radius. The total flux of the disk is measured when the COG reaches a peak in emission. For fainter sources, such as USco 4, negative bowls in the reconstructed images can lead to multiple peaks in the COG. In these cases, the first peak is selected. The flux measurements for the continuum, $^{12}$CO, $^{13}$CO, $C^{18}$O, and $N_2H^+$ obtained from the COG are detailed in Table 4. In

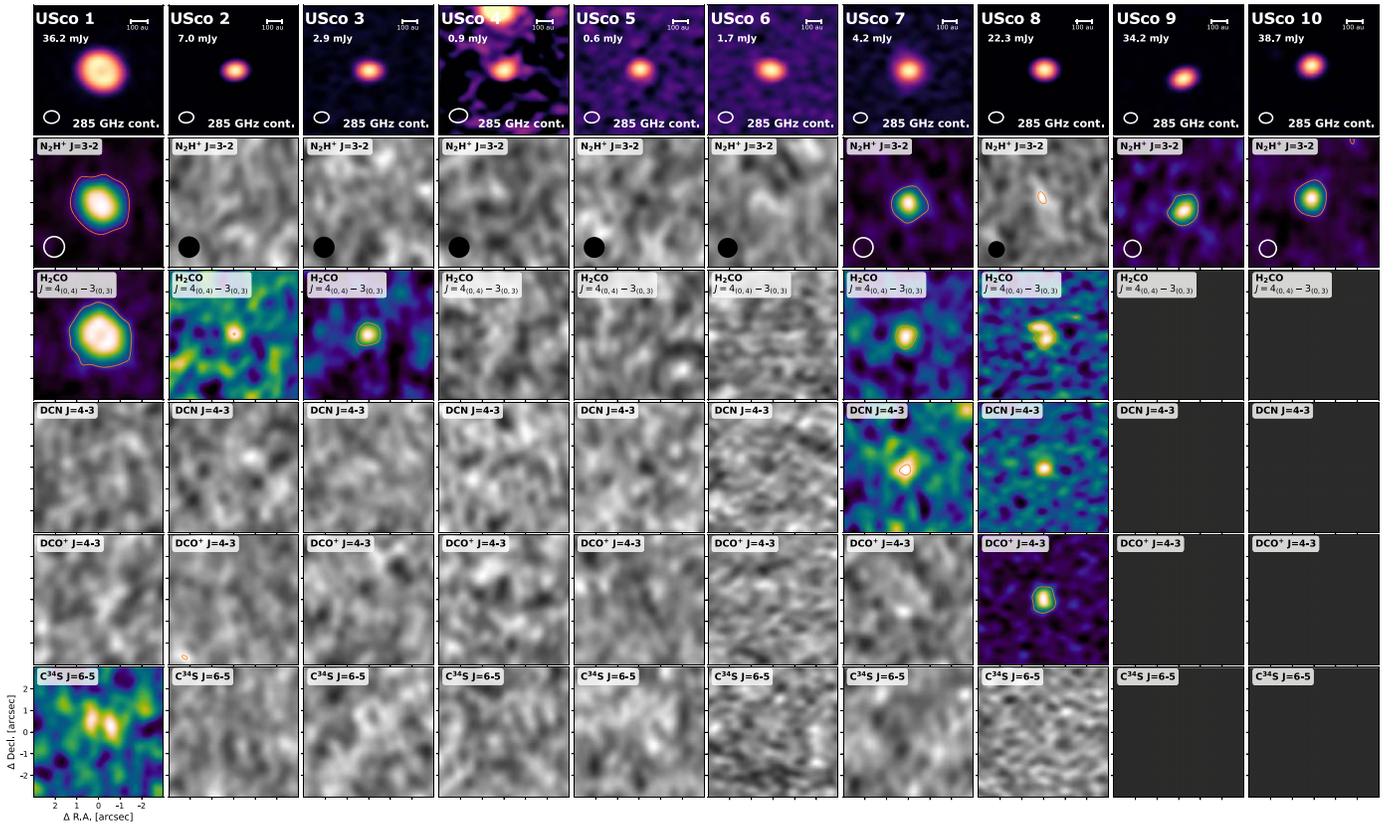

**Figure 8.** Gallery of Band 7 dust continuum images and line moment zero maps for our Upper Sco sample. Disks with clear detections are shown in color. Nondetections are shown in grayscale. Lines not included in the AGE-PRO observational setup are dark gray. The northern source within the USco 4 dust map has been identified as a background galaxy.

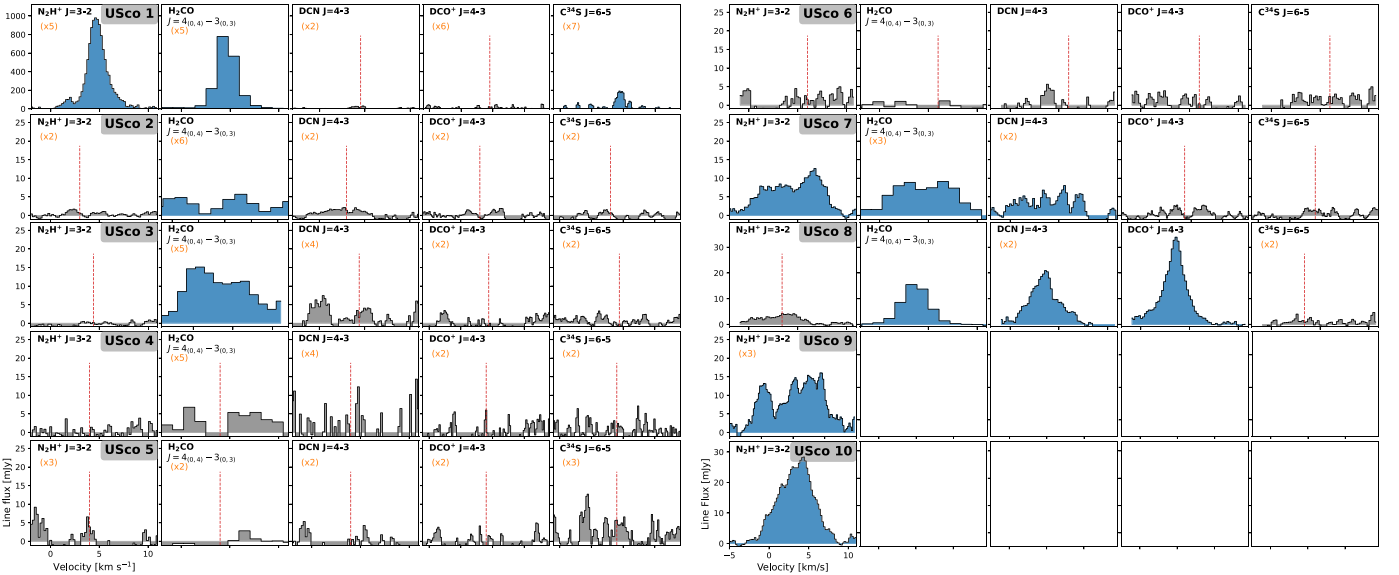

**Figure 9.** Disk-integrated line spectra of $N_2H^+$, $H_2CO$, DCN, $DCO^+$, and $C^{34}S$ for Upper Sco targets in Band 7. We use the aperture corresponding to the gas disk radii for $^{12}CO(J = 2–1)$. Disks without a $3\sigma$ detection of line emission are shown in gray, with red dashed lines indicating the systemic velocity identified from the $^{12}CO$ line.

Appendix C, we compare our flux measurements in Bands 6 and 7 for both continuum and molecular lines with existing literature flux values (S. A. Barenfeld et al. 2016; D. E. Anderson et al. 2019; J. B. Bergner et al. 2020; J. Pegues et al. 2020), confirming that our Upper Sco measurements are in agreement with previously published studies. We corroborate that the COG method works well even for unresolved or marginally resolved observations, since the peak of emission in these disks is consistent with the total flux obtained with the COG method.

Additionally, for both continuum and $^{12}CO$ emission, we determine $R_{68}$ and $R_{90}$ as the radius enclosing 68% and 90% of the peak flux, respectively. Uncertainties in these radii are estimated by calculating the range of radius corresponding to the uncertainties in the 68% and 90% flux levels.

Table 4
Measured Fluxes from the COG method

| ID | $F_{226\,\text{GHz}}$ (mJy) | $F_{285\,\text{GHz}}$ (mJy) | $F_{^{12}\text{CO}(2-1)}$ (mJy km s$^{-1}$) | $F_{^{13}\text{CO}(2-1)}$ (mJy km s$^{-1}$) | $F_{\text{C}^{18}\text{O}(2-1)}$ (mJy km s$^{-1}$) | $F_{\text{N}_2\text{H}^+(3-2)}$ (mJy km s$^{-1}$) |
|---|---|---|---|---|---|---|
| USco 1  | 16.1 ± 0.1    | 34.3 ± 0.1   | 3171.6 ± 11  | 781.8 ± 8   | 151.3 ± 5  | 444.8 ± 7 |
| USco 2  | 4.3 ± 0.1     | 7.03 ± 0.03  | 97.6 ± 5     | 31.4 ± 10   | 18.4 ± 5   | 1.6 ± 2 |
| USco 3  | 1.6 ± 0.1     | 2.99 ± 0.04  | 312.8 ± 5    | 101.3 ± 3   | 39.0 ± 2   | ⋯ |
| USco 4  | 0.13 ± 0.02   | 0.21 ± 0.02  | 222.0 ± 5    | 18.7 ± 4    | 12.7 ± 4   | ⋯ |
| USco 5  | 0.57 ± 0.04   | 0.68 ± 0.03  | 275.4 ± 7    | 10.5 ± 4    | ⋯          | ⋯ |
| USco 6  | 0.98 ± 0.03   | 1.91 ± 0.03  | 201.0 ± 17   | ⋯           | ⋯          | 7.8 ± 3 |
| USco 7  | 1.6 ± 0.1     | 3.14 ± 0.03  | 1506.3 ± 15  | 180.3 ± 10  | 28.4 ± 8   | 95.2 ± 5 |
| USco 8  | 12.6 ± 0.2    | 22.2 ± 0.1   | 1917.9 ± 9   | 275.4 ± 7   | 63.0 ± 4   | 27.3 ± 4 |
| USco 9  | 16.30 ± 0.02  | 34.9 ± 0.1   | 481.3 ± 6    | 157.5 ± 10  | 114.2 ± 7  | 112.4 ± 4 |
| USco 10 | 22.3 ± 0.1    | 38.0 ± 0.1   | 630.8 ± 13   | 180.0 ± 5   | 52.0 ± 17  | 180.1 ± 5 |

Table 5
Measured Dust and Gas Radii from the COG Method

| ID | Continuum at 226 GHz | | | | | $^{12}$CO (2–1) | | | | |
|---|---|---|---|---|---|---|---|---|---|---|
|  | Beam (arcsec) | $R_{68}$ (arcsec) | (au) | $R_{90}$ (arcsec) | (au) | Beam (arcsec) | $R_{68}$ (arcsec) | (au) | $R_{90}$ (arcsec) | (au) |
| USco 1  | 0.27 × 0.27 | 0.711 ± 0.002 | 93.8 | 0.84 ± 0.01   | 110.3 | 0.27 × 0.27 | 0.994 ± 0.003 | 144.1 | 1.346 ± 0.007 | 177.8 |
| USco 2  | 0.32 × 0.32 | 0.329 ± 0.001 | 45.3 | 0.45 ± 0.01   | 62.5  | 0.33 × 0.33 | 0.35 ± 0.02   | 50.6  | 0.49 ± 0.05   | 70.9 |
| USco 3  | 0.35 × 0.35 | 0.449 ± 0.005 | 62.7 | 0.64 ± 0.01   | 88.8  | 0.35 × 0.35 | 0.542 ± 0.008 | 76.3  | 0.82 ± 0.03   | 115.9 |
| USco 4  | 0.34 × 0.34 | 0.51 ± 0.83   | 69.4 | 0.69 ± 1.34   | 94.3  | 0.34 × 0.34 | 0.64 ± 0.02   | 88.7  | 0.98 ± 0.06   | 134.5 |
| USco 5  | 0.29 × 0.29 | 0.27 ± 0.02   | 38.9 | 0.36 ± 0.06   | 52.0  | 0.29 × 0.29 | 0.320 ± 0.006 | 46.4  | 0.44 ± 0.02   | 63.5 |
| USco 6  | 0.27 × 0.27 | 0.40 ± 0.01   | 64.0 | 0.53 ± 0.04   | 84.1  | 0.29 × 0.29 | 0.52 ± 0.05   | 75.8  | 1.1 ± 0.4     | 160.2 |
| USco 7  | 0.29 × 0.29 | 0.65 ± 0.07   | 99.3 | 0.97 ± 0.07   | 147.5 | 0.32 × 0.32 | 0.940 ± 0.008 | 144.8 | 1.24 ± 0.02   | 180.2 |
| USco 8  | 0.38 × 0.38 | 0.321 ± 0.004 | 44.6 | 0.43 ± 0.01   | 59.6  | 0.39 × 0.39 | 0.932 ± 0.004 | 129.6 | 1.35 ± 0.01   | 188.2 |
| USco 9  | 0.33 × 0.33 | 0.566 ± 0.001 | 77.5 | 0.832 ± 0.002 | 114.0 | 0.35 × 0.35 | 0.668 ± 0.009 | 91.5  | 1.02 ± 0.05   | 139.3 |
| USco 10 | 0.31 × 0.31 | 0.370 ± 0.001 | 51.2 | 0.523 ± 0.002 | 71.6  | 0.33 × 0.33 | 0.461 ± 0.009 | 63.4  | 0.64 ± 0.03   | 87.8 |

The associated disk radii for both dust and gas tracers are summarized in Table 5.

### 4.5. Spectral Indices and Dust Masses

Dust continuum emission at millimeter wavelengths can constrain the presence of millimeter-sized dust grains (see, e.g., L. Testi et al. 2014). At these wavelengths, the observed dust emission can be approximated as $F_\nu \propto \kappa_\nu B_\nu(T_d)\Sigma$, with $\kappa_\nu$ the dust opacity, $T_d$ the dust temperature, and $\Sigma$ the surface density of dust. The dust opacity can also be approximated as $\kappa_\nu \propto \nu^{\beta_{\text{mm}}}$, where $\beta_{\text{mm}}$ is the dust opacity spectral index. For optically thin emission in the Rayleigh–Jeans regime, the emission spectral index is simply $\alpha_{\text{mm}} = \beta_{mm} + 2$, where $F_\nu \propto \nu^{\alpha_{\text{mm}}}$. Using the continuum fluxes from our Band 6 and 7 observations at 226 and 285 GHz, respectively, we compute the global (disk-averaged) millimeter spectral index, $\alpha_{\text{mm}}$, presented for each source in Table 6. Uncertainties in $\alpha_{\text{mm}}$ are estimated following the procedure in H.-F. Chiang et al. (2012), where the uncertainty in $\alpha$ is propagated assuming that the absolute flux error can be represented as a Gaussian noise with a standard deviation of 10%.

We estimate a disk solid mass from the Band 6 (1.3 mm) continuum fluxes for each disk. We adopt the same approach as in S. M. Andrews et al. (2013), assuming the dust continuum fluxes are optically thin and derive dust disk masses from the equation

$$M_{\text{dust}} = \frac{d^2}{B_\nu(T_{\text{dust}})\kappa_\nu}F_\nu, \quad (1)$$

where $d$ is the distance to the object, $B_\nu(T_{\text{dust}})$ is the Plank spectrum at the average dust temperature $T_{\text{dust}}$, and $\kappa_\nu$ is the dust opacity. We adopt an average dust temperature of 20 K and a dust opacity $\kappa_\nu = 2.3\,(\nu/230\,\text{GHz})$ [cm$^2$ g$^{-1}$]. The derived dust disk masses are presented in Table 6.

### 4.6. Sources Warranting Further Investigation

Within the Upper Sco sample, there are two objects of individual scientific interest, with previously unknown peculiar aspects.

#### 4.6.1. Potential Planet Formation Signatures in USco 1

The analysis of the protoplanetary disk around USco 1, presented in A. Sierra et al. (2024), revealed several features consistent with ongoing planet formation in this system. In dust continuum emission, USco 1 appears as a wide bright ring of ~100 au radius with an unresolved faint inner disk. Between the bright ring and faint inner disk, there is a compact point source at the center of the disk gap, which appears to connect with a spiral arm structure traced after subtracting the azimuthally averaged disk emission. In $^{12}$CO emission, the disk extends to ~200 au, and its velocity field is well described by a Keplerian model with an elevated disk surface that increases with radius. Subtraction of the best-fit Keplerian rotation model shows in the residuals a localized deviation in the gas velocity structure with respect to the Keplerian rotation, located in the outer edge of the disk continuum emission. These features, observed in both gas and dust, are studied in detail by A. Sierra et al. (2024), since they might represent signatures of planet formation.

**Table 6**
Millimeter Spectral Index and Inferred Dust Mass

| ID | $\alpha_{mm}$ | $M_{dust}$ ($M_\oplus$) |
|---|---|---|
| 1 | 3.2 ± 0.6 | 9.0 |
| 2 | 2.1 ± 0.6 | 2.6 |
| 3 | 2.6 ± 0.6 | 1.0 |
| 4 | 2.0 ± 1.0 | 0.1 |
| 5 | 1.9 ± 0.7 | 0.4 |
| 6 | 2.5 ± 1.1 | 0.9 |
| 7 | 3.1 ± 0.7 | 1.2 |
| 8 | 2.4 ± 0.6 | 7.8 |
| 9 | 3.2 ± 0.6 | 9.8 |
| 10 | 2.2 ± 0.6 | 13.4 |

### 4.6.2. A Flare in USco 7

We observed a significant variation in the continuum flux density toward USco 7 during the SB observations. The first of the three SB executions is 10 times brighter than the disk emission in quiescent levels, as measured in the other SB and LB executions. The brightness of the flare emission decays to quiescent levels during the ≈1 hr observation, suggesting that the phenomenon is short lived and that the true peak brightness remains unconstrained. USco 7 had previously been considered typical, classified as an M2 type star with a mass of 0.5 $M_\odot$ (C. F. Manara et al. 2023). Archival ALMA observations indicate a faint disk in both molecular emission and the dust continuum, with no evidence for outflows or envelopes. This unusual behavior is analyzed in a separate follow-up study by J. M. Miley et al. (2025).

## 5. Discussion

In this section, we discuss general trends for the inferred properties of the Upper Sco disks targeted in AGE-PRO. First, evidence for radial drift and dust trapping, based on the measured gas and dust disk sizes, is presented in Section 5.1. Next, a discussion related to different findings on the solids of Upper Sco is presented in Section 5.2. Then, possible correlations between dust and gas emission in Upper Sco are discussed in Section 5.3. Finally, a comparison with the younger Lupus disks of AGE-PRO is presented in Section 5.4.

### 5.1. Gas-dust Disk Size and Implications for Radial Drift

The ratio between the gas and dust disk radius can be used to identify sources with significant dust evolution and radial drift (e.g., L. Trapman et al. 2019, 2020; C. Toci et al. 2021). Figure 10 compares the $^{12}$CO radius and the dust continuum radius in Band 6 at 226 GHz, at both 68% and 90% of their total flux. These radii trace the disk size in gas ($R_{gas}$) and the disk size in dust ($R_{dust}$), respectively. Both tracers have beam sizes of ∼0″.35 and are represented by vertical and horizontal green dashed lines in each panel of Figure 10.

For USco 1, which is well resolved in the images of both the $^{12}$CO (2–1) and Band 6 continuum, the CO gas radius is 1.4 and 1.6 times larger than the dust radius (for both $R_{68}$ and $R_{90}$, respectively). However, for most Upper Sco systems, we constrain dust disk radii that are similar to the beam size. Even so, Figure 10 shows that all Upper Sco disks lie outside the drift-dominated region, indicating empirically that a degree of efficient drift halting exists. This finding is consistent with the results of M. Vioque et al. (2025), who performed visibility fitting of the Band 6 continuum observations to obtain deconvolved dust radial profiles and better infer the dust disk radii. A comparison between dust and gas disk radii as measured from the image plane (this work) and measured from deconvolution methods (L. Trapman et al. 2025b; M. Vioque et al. 2025) is presented in Figure D1 in Appendix D.

### 5.2. Solids in Upper Sco

Dust masses constrained in our AGE-PRO Upper Sco sample range from 0.08 to 13 $M_\oplus$, while in AGE-PRO Lupus, the dust masses range from 1.2 to 166 $M_\oplus$ (D. Deng et al. 2025b), indicating a lower dust budget in the older Upper Sco region of the AGE-PRO survey. Regarding the millimeter dust continuum spectral index, values around $\alpha_{mm} \sim 3.5$ are generally associated with the sizes of interstellar medium particles (A. Natta et al. 2007; L. Testi et al. 2014), while material in protoplanetary disks shows clear signs of dust growth, with a lower spectral index of $\alpha_{mm} \lesssim 3$ (L. Testi et al. 2014, and references therein). In our AGE-PRO Upper Sco sample, we find spectral indices that agree with larger dust grains ($\alpha_{mm} \lesssim 3$), and which are as low as expected for optically thick emission ($\alpha_{mm} \approx 2$). The relation of dust mass and spectral index with observational properties, such as disk gas mass, stellar mass, and gas-to-dust mass ratio, are further discussed in N. T. Kurtovic et al. (2025).

Several findings from the AGE-PRO survey support the idea that a combination of solids growing into larger bodies, radial drift, and dust trapping is taking place in these older disks: (1) the median dust disk mass in Upper Sco is only 1.8 $M_\oplus$, which is significantly lower compared to the younger Ophiuchus and Lupus AGE-PRO samples (D. Deng et al. 2025b; D. A. Ruiz-Rodriguez et al. 2025); (2) the median dust disk size remains consistent across regions (K. Zhang et al. 2025), and when compared to the gas disk size, Upper Sco disks are outside of the drift-dominated region (see Section 5.1); and (3) the millimeter spectral index increases from Lupus to Upper Sco (K. Zhang et al. 2025). The lower dust masses can be attributed to the combined effect of dust grains growing into larger bodies and their subsequent inward drift (e.g., J. Appelgren et al. 2023), a combination that may also be responsible for the increased spectral index observed in Upper Sco. And coupled with the dust disk size being similar from young to older ages, and not consistent with a drift-dominated regime, it indicates that some grains must remain trapped at large disk radii, preventing the solids from completely disappearing from the outer regions of the disk (e.g., P. Pinilla et al. 2020). In N. T. Kurtovic et al. (2025), theoretical models of dust drift with some degree of trapping are able to reproduce these findings.

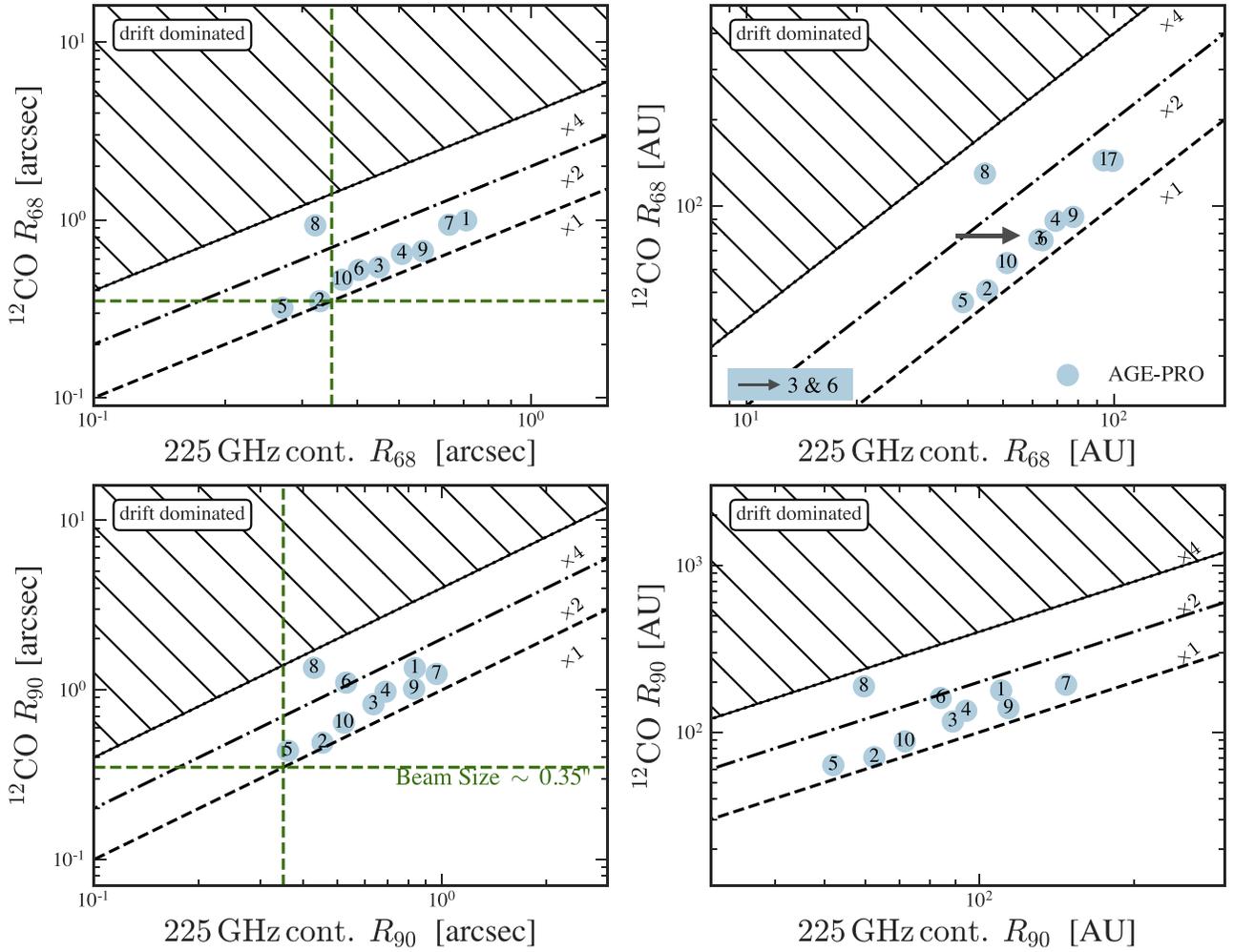

**Figure 10.** Gas and dust disk sizes, traced by the disk radius of $^{12}$CO line emission and Band 6 dust continuum emission. Upper and bottom panels show $R_{68}$ and $R_{90}$, respectively, while right and left panels show disk radii in units of au and arcsec, respectively. Typical beam sizes are shown in green dashed vertical and horizontal lines. The three dashed lines show where the $^{12}$CO radius is 1 times, 2 times, and 4 times the dust disk radius. The shaded region corresponds to radial drift-dominated disks, where the gas disk size is over 4 times larger than the dust disk size (L. Trapman et al. 2019, 2020). The radii reported here are circularized by using the CASA task `imsmooth`. The deconvolved radius measurements are reported in separate publications (L. Trapman et al. 2025b; M. Vioque et al. 2025).

### 5.3. Correlations Between Dust and Gas Emission

Prior shallow surveys of nearby star-forming regions (e.g., M. Ansdell et al. 2016; S. A. Barenfeld et al. 2016; L. A. Cieza et al. 2019) typically detect dust continuum more readily than gas lines, particularly less abundant molecules. AGE-PRO's higher sensitivity unlocks a wealth of detections for rarer CO isotopologues, allowing us to directly probe the correlation between molecular line fluxes and continuum emission at different ages. Direct empirical correlations could inform future gas observations, aiding in disk and planet formation studies. In addition to CO isotopologues, AGE-PRO also provides $N_2H^+(J=3–2)$ line detections, which can be used to measure the CO abundance in the disk together with $C^{18}O(J=2–1)$ (e.g., L. Trapman et al. 2022). Thus, we also empirically compare the $N_2H^+(J=3–2)$ fluxes with the $C^{18}O(J=2–1)$ fluxes.

Figure 11 shows the comparison between CO and its isotopologues' line fluxes with Band 6 continuum fluxes, as well as the $C^{18}O(J=2–1)$ versus $N_2H^+(J=3–2)$ line emission fluxes. From the figure, it seems that CO isotopologue fluxes are not tightly correlated with the $F_{226\,GHz\,cont}$ continuum emission, while the $C^{18}O(J=2–1)$ line flux, $F_{C^{18}O(J=2–1)}$, appears to be better correlated with the $N_2H^+(J=3–2)$ line emission, $F_{N_2H^+(J=3–2)}$.

We employ the `pymccorrelation` routine v0.2.5[25] to test the significance of these empirical correlations. The routine carries out the nonparametric Kendall's $\tau$ test with censored data (T. Isobe et al. 1986) and is capable of including uncertainties and upper limits. The resulting median values of Kendall's $\tau$ and the percent probability $p$ are presented in each panel of Figure 11, together with the best-fit linear regression for the data. In this test, $\tau$ gives the direction of the correlation (positive correlation for values $\tau > 0$), while $p$ is the percent probability that the quantities are not correlated. We find that all three CO isotopologue fluxes are positively but not tightly correlated with the disk continuum emission (positive $\tau$ and $p > 4\%$). Similarly, $F_{C^{18}O(2–1)}$ and $F_{N_2H^+(3–2)}$ might be positively correlated, since $p \sim 4\%$. We compare the observed correlation between gas lines and dust fluxes in Upper Sco with those in Lupus (from D. Deng et al. 2025b), displaying in each panel of Figure 11 the best-fit linear regression for Lupus as a dashed line and the $p$ probability value from the Kendall's $\tau$ test. In Section 5.4, we further discuss and compare these possible empirical correlations in these two regions of different age.

---

[25] https://github.com/privong/pymccorrelation

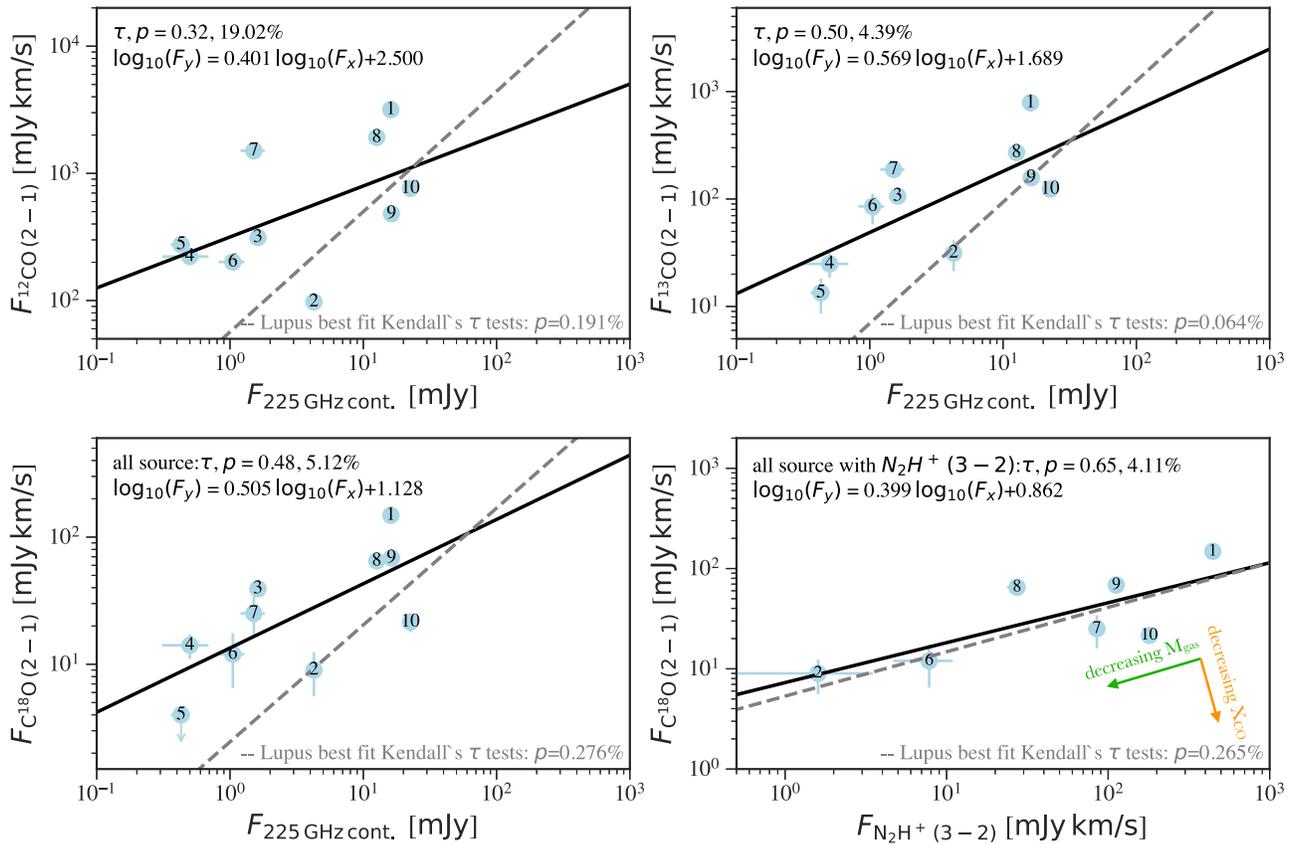

**Figure 11.** Comparison between fluxes of gas and dust tracers for the AGE-PRO sample of disks in Upper Sco. The results of pymccorrelation Kendall's $\tau$ tests for the Upper Sco sample are reported in each panel, suggesting a positive correlation with large scatter between gas and dust emission. The best-fit results from linear regression are plotted in black solid lines and include all sources, except for $N_2H^+(J=3-2)$ in which USco 3, 4, and 5 have nondetections. The gray dashed line plotted on each panel is the best-fit linear regression for the Lupus disk sample from D. Deng et al. (2025b), with the Kendall's $p$-value for Lupus in the bottom-right corner of each panel.

In both the $^{13}$CO and C$^{18}$O fluxes versus dust continuum fluxes (top-right and bottom-left panels of Figure 11), it seems that USco 2 and 10 are outliers, with lower gas line fluxes for both $^{13}$CO and C$^{18}$O, but relatively large $F_{226\,\text{GHz cont.}}$ compared to disks with similar $F_{^{13}\text{CO}(2-1)}$ and $F_{\text{C}^{18}\text{O}(2-1)}$ (e.g., USco 4, 6, and 7). If USco 2 and 10 are removed from the analysis, the correlation between $F_{\text{C}^{18}\text{O}(2-1)}$ and $F_{226\,\text{GHz cont.}}$ becomes much stronger ($p = 0.45\%$), with a similar trend and slightly steeper slope.

Upon assumption of optically thin emission, $F_{\text{C}^{18}\text{O}(2-1)}$ and $F_{226\,\text{GHz cont.}}$ trace the column densities of gas and dust, respectively, and thus their observed positive correlation suggests that across the majority of the sample the CO gas-to-dust mass ratio is similar. However, USco 2 and 10 deviate from this trend and are relatively younger than the other Upper Sco systems (see Table 1), potentially indicating variations in their individual gas-to-dust ratios. Another option is that CO is more depleted in these systems (and in USco 5) than the rest, which would result in lower abundances of C$^{18}$O and thus lower $F_{\text{C}^{18}\text{O}(2-1)}$ line flux, as observed. This possibility is compatible with USco 10, since it falls below the correlation of $F_{\text{C}^{18}\text{O}(2-1)}$ with both $F_{226\,\text{GHz cont.}}$ and $F_{N_2H^+(3-2)}$ (bottom panels of Figure 11).

It has been proposed that combined analysis of the C$^{18}$O (2–1) and $N_2H^+$ (3–2) line fluxes can help break the degeneracy between the gas disk mass and CO abundance ($X_{\text{CO}}$). This approach takes advantage of the fact that the main formation and destruction paths of $N_2H^+$ involve reactions with gaseous CO that can deplete $N_2H^+$, as discussed in detail by, for example, D. E. Anderson et al. (2019). L. Trapman et al. (2022) demonstrated that the relative trends between $F_{N_2H^+(3-2)}$ and $F_{\text{C}^{18}\text{O}(2-1)}$ depend on the abundance of CO within the warm molecular layer, specifically, the degree of CO depletion compared to its abundance of the interstellar medium. When $X_{\text{CO}}$ decreases due to destruction, a negative correlation between the two fluxes is expected. In contrast, a positive correlation might indicate a decrease in $M_{\text{gas}}$ without significant CO depletion. As seen in the bottom-right panel of Figure 7, these two trends are indicated by the orange and green arrows (for a decrease in $X_{\text{CO}}$ and a decrease in $M_{\text{gas}}$, respectively). In the AGE-PRO Upper Sco sample, we find a positive correlation between the line emission of C$^{18}$O (2–1) and $N_2H^+$ (3–2), which empirically suggests that the level of CO depletion is similar between the AGE-PRO Upper Sco disks (more details are in L. Trapman et al. 2025b).

### 5.4. Comparison with the Younger Lupus Region

Interestingly, Figure 11 reveals a disparity between the correlation slopes constrained for Lupus (as derived by D. Deng et al. 2025b) and for Upper Sco (this work) when comparing gas and dust tracers (e.g., $F_{\text{C}^{18}\text{O}(2-1)}$ versus $F_{226\,\text{GHz cont.}}$). For similar C$^{18}$O line fluxes, the younger disks in Lupus exhibit greater dust emission than the older Upper Sco disks. We interpret this as an empirical indication that dust masses may evolve faster than gas masses, although our interpretation is limited given the large scatter in gas versus dust fluxes for our Upper Sco sample, and the larger Kendall's $p$ probability for these correlations ($p > 4\%$ in Upper Sco) when compared to Lupus ($p < 0.3\%$). However, L. Trapman et al. (2025b) find a similar evolutionary trend when more sophisticated modeling is used to estimate gas masses. One possibility to explain the larger scatter in Upper Sco when compared to Lupus is that disks in the Upper Sco region have a larger age spread (S. Ratzenböck et al. 2023), and thus not the same evolutionary state. For a detailed discussion about the AGE-PRO ages and the young and old subgroups within Upper Sco, please refer to K. Zhang et al. (2025). Another option is that individual disks have different CO abundances, independent from each other. Nevertheless, the larger scatter observed in the correlation between gas and dust content might be due to a combination of these mechanisms or further unknown ones (R. Anania et al. 2025; B. Tabone et al. 2025).

Remarkably, the empirical correlation between $F_{C^{18}O\,(2-1)}$ and $F_{N_2H^+\,(3-2)}$ (bottom-right panel of Figure 11) exhibits an indistinguishable linear trend for the older Upper Sco disks (continuous line) and the younger Lupus systems (dashed line). If the CO abundance, $X_{CO}$, evolves over time, we would expect an offset between these two trends. However, we find that $X_{CO}$ remains similar to that of Lupus for the more evolved Upper Sco region, consistent with the AGE-PRO modeling results in L. Trapman et al. (2025b).

Another aspect to highlight is the fact that the dust disk sizes for the Upper Sco systems are similar to or larger than the dust disk sizes in Lupus (D. Deng et al. 2025b). The presence of dust traps, which halt the expected inward drift of particles, could explain why the solids in these older Upper Sco disks still remain at large radii. This is consistent with previous studies that anticipate dust traps at large radii in older disks in order to explain observed disk sizes (P. Pinilla et al. 2020; N. van der Marel & G. D. Mulders 2021). Even with the current spatial resolution of our AGE-PRO observations, M. Vioque et al. (2025) have constrained the presence of large inner dust cavities in some Upper Sco systems, which is consistent with the dust trap scenario. The continuum visibility fitting analysis of M. Vioque et al. (2025) reveals a higher incidence of large inner dust cavities in the Upper Sco sample (four out of the six well-resolved disks) compared to younger AGE-PRO populations. Furthermore, the dust evolution modeling of N. T. Kurtovic et al. (2025) suggests that the dust content of the AGE-PRO disks in Upper Sco has survived until their current age due to the presence of, at least weak, dust traps. This aligns with the evolutionary sequence proposed by M. Vioque et al. (2025), where the surviving disk population in Upper Sco has a higher fraction of disks with dust traps because these have prevented the inward drift and dissipation of their dust disks. This causes a larger fraction of disks with large inner dust cavities in older populations. This would explain the observed high fraction of large inner dust cavities in Upper Sco, the apparent nonevolution of $R_{68\%}$ over time, and the absence of the $R_{dust}$–$L_{mm}$ correlation (see M. Vioque et al. 2025). We acknowledge the limitations of the current AGE-PRO observations in spatially resolving these dust traps. The current spatial resolution is not sufficient to definitively confirm their presence. Future multiwavelength and spatially resolved observations are crucial for discerning the presence and nature of these traps. These additional data will enable more detailed investigations of the interplay between dust traps, grain growth, and the resulting dust disk sizes in Upper Sco and other star-forming regions.

## 6. Summary

This study presents a comprehensive ALMA survey of 10 protoplanetary disks of the Upper Sco region, the oldest observed within the AGE-PRO program. These are currently the most sensitive observations of disks in this region that simultaneously probe dust continuum emission, $^{12}$CO, CO isotopologues, and $N_2H^+$ emission lines. Our analysis reveals intriguing trends, potentially linked to the evolved nature of Upper Sco disks compared to younger regions. Our conclusions are as follows:

1. We report detections of dust continuum and $^{12}$CO line emission in all 10 sources, while CO isotopologues and $N_2H^+$ were primarily detected in continuum-bright disks, with the notable exception of USco 2.
2. We constrain a more evolved solid component in the older Upper Sco sample, since the dust masses are, on average, an order of magnitude smaller than in the younger regions probed by AGE-PRO, with millimeter spectral index values consistent with evolved larger grains.
3. The dust disk sizes of the Upper Sco targets in the image plane are comparable to or even larger than their Lupus counterparts. This difference might be explained by the presence of further-out dust traps within these older disks, which hinders the expected inward drift of dust grains.
4. An empirical correlation between less abundant CO isotopologues and continuum fluxes is observed, suggesting a connection between the gas and dust content. However, the scatter in this relation is considerably larger compared to that found in the younger Lupus region. This discrepancy could be attributed to several factors: the presence of disks with varying ages within Upper Sco, or the possibility of different CO abundances within individual disks, or a combination of both, or even other unknown mechanisms.
5. The empirical correlation observed between $C^{18}O$ and $N_2H^+$ line fluxes remains unchanged for the younger Lupus disks and the older Upper Sco disks, evidence that the CO gas abundance remains nearly constant in the Class II stages of disk evolution.
6. We measure the 68% and 90% radius of Band 6 dust continuum and $^{12}$CO emission using the COG method in the image plane. Only one disk is well resolved, USco 1, which shows a ratio between the gas and dust disk size of $\sim$1.5, which indicates that there is no significant radial drift. For unresolved targets, gas and dust radii are derived using Nuker profiles and visibility fitting, as detailed in L. Trapman et al. (2025a) and M. Vioque et al. (2025), respectively. These disks have sizes in gas and dust that, overall, point toward radial drift being halted in the AGE-PRO Upper Sco sample.

## Acknowledgments

This paper makes use of the following ALMA data: ADS/JAO.ALMA#2021.1.00128.L. ALMA is a partnership of ESO (representing its member states), NSF (USA) and NINS (Japan), together with NRC (Canada), MOST and ASIAA (Taiwan), and KASI (Republic of Korea), in cooperation with the Republic of Chile. The Joint ALMA Observatory is operated by ESO, AUI/NRAO and NAOJ. The National Radio Astronomy Observatory is a facility of the National Science Foundation operated under cooperative agreement by Associated Universities, Inc.

C.A.G. acknowledges support from FONDECYT de Postdoctorado 2021 \#3210520. C.A.G. and L.P acknowledge support from "Comité Mixto ESO-Gobierno de Chile 2023" under grant 072-2023. A.S. acknowledges support from FONDECYT de Postdoctorado 2022 \#3220495 and support from the UK Research and Innovation (UKRI) under the UK government's Horizon Europe funding guarantee from ERC (under grant agreement No. 101076489). J.M. acknowledges support from FONDECYT de Postdoctorado 2024 #3240612; J.M. acknowledges support from the Millennium Nucleus on Young Exoplanets and their Moons (YEMS), ANID—Center Code NCN2024_001, and L.A.C. also acknowledges support from the FONDECYT grant #1241056. N.T.K. acknowledges support provided by the Alexander von Humboldt Foundation in the framework of the Sofja Kovalevskaja Award endowed by the Federal Ministry of Education and Research. G.R. and R.A. acknowledge funding from the Fondazione Cariplo, grant No. 2022-1217, and the European Research Council (ERC) under the European Union's Horizon Europe Research and Innovation Programme under grant agreement No. 101039651 (DiscEvol). Views and opinions expressed are, however, those of the author(s) only, and do not necessarily reflect those of the European Union or the European Research Council Executive Agency. Neither the European Union nor the granting authority can be held responsible for them. K.S. acknowledges support from the European Research Council under the Horizon 2020 Framework Program via the ERC Advanced grant Origins 83 24 28. B.T. acknowledges support from the Programme National "Physique et Chimie du Milieu Interstellaire" (PCMI) of CNRS/INSU with INC/INP and cofunded by CNES.

All figures were generated with the PYTHON-based package MATPLOTLIB (J. D. Hunter 2007). This research made use of Astropy,[26] a community-developed core Python package for Astronomy (Astropy Collaboration et al. 2013, 2018), and Scipy (P. Virtanen et al. 2020).

## Appendix A
## Observation Log for AGE-PRO Upper SCO Sample

The observational log and UIDs for the executions of the AGE-PRO Upper Sco observations, as well as the archival data, are summarized in Tables 7 and 8, respectively.

Table 7
ALMA Observational Log for AGE-PRO Upper Sco Sample

| Setup | USco ID | UTC Date | Config | Baselines (m) | $N_{ant}$ | Elev (deg) | PWV (mm) | Calibrators |
|---|---|---|---|---|---|---|---|---|
| USco CO | 2, 3, 4, 9, 10 | 2022-07-03 0:08:57 | C43-5 | 15–1301 | 42 | 73.3 | 0.4 | J1517-2422, J1617-1941 |
| | 2, 3, 4, 9, 10 | 2022-07-02 02:19:07 | C43-5 | 21–1301 | 45 | 72.9 | 0.4 | J1517-2422, J1553-2422 |
| | 2, 3, 4, 9, 10 | 2022-07-01 23:20:38 | C43-5 | 15–1301 | 46 | 64.0 | 0.4 | J1517-2422, J1553-2422 |
| | 2, 3, 4, 9, 10 | 2022-07-01 02:20:45 | C43-5 | 15–1301 | 46 | 73.3 | 0.5 | J1517-2422, J1553-2422 |
| | 2, 3, 4, 9, 10 | 2022-03-28 08:36:51 | C43-2 | 14–313 | 45 | 76.9 | 1.6 | J1517-2422, J1553-2422 |
| | 2, 3, 4, 9, 10 | 2022-03-29 09:30:24 | C43-2 | 14–313 | 45 | 63.4 | 1.8 | J1517-2422, J1553-2422 |
| | 1, 5, 6, 7, 8 | 2022-07-03 23:08:00 | C43-5 | 15–1301 | 42 | 61.0 | 1.1 | J1517-2422, J1626-2951 |
| | 1, 5, 6, 7, 8 | 2022-07-04 01:05:51 | C43-5 | 15 - 1301 | 41 | 84.2 | 1.1 | J1427-4206, J1626-2951 |
| | 1, 5, 6, 7, 8 | 2022-07-04 02:42:03 | C43-5 | 15 - 1301 | 41 | 68.7 | 1.2 | J1517-2422, J1626-2951 |
| | 1, 5, 6, 7, 8 | 2022-07-05 00:39:42 | C43-5 | 15 - 1996 | 41 | 82.7 | 1.8 | J1427-4206, J1626-2951 |
| | 1, 5, 6, 7, 8 | 2022-07-17 23:38:35 | C43-5 | 15–2617 | 44 | 81.1 | 0.4 | J1427-4206, J1700-2610 |
| | 1, 5, 6, 7, 8 | 2022-03-30 08:27:26 | C43-2 | 14–313 | 46 | 79.1 | 0.9 | J1517-2422, J1625-2527 |
| | 1, 5, 6, 7, 8 | 2022-04-07 05:49:46 | C43-2 | 14–313 | 46 | 70.9 | 2.6 | J1517-2422, J1625-2527 |
| USco $N_2H^+$ | 9, 10 | 2016 March 9:35:00 | C43-2 | 15.2–460 | 41 | 78.3 | 1.4 | J1517-2422, J1626-2951, J1733-1304 and Titan |
| | 9, 10 | 2016 March 8:34:00 | C43-2 | 15.1–460 | 40 | 72.1 | 1.4 | J1517-2422, J1553-2422, J1733-1304 and Titan |
| | 9, 10 | 2016 March 11:58:00 | C43-2 | 15.1–460 | 39 | 48.7 | 1.4 | J1517-2422, J1553-2422, J1733-1304 and Titan |
| | 9, 10 | 2016 March 11:39:00 | C43-2 | 15.1–460 | 38 | 52.2 | 1.4 | J1517-2422, J1553-2422, J1733-1304 and Titan |
| | 1 | 2022-04-12 4:32 | C43-3 | 15.1–455.5 | 42 | 59.4 | 0.6 | J1517-2422 and J1626-2951 |
| | 1 | 2022-04-12 5:34 | C43-3 | 15.1–455.5 | 42 | 72.5 | 0.6 | J1517-2422 and J1626-2951 |
| | 2–8 | 2022-04-17 5:14 | C43-3 | 15.1–500.2 | 46 | 75.1 | 0.3 | J1517-2422 and J1553-2422 |
| | 2–8 | 2022-04-22 8:51 | C43-3 | 15.1–500.2 | 44 | 50.1 | 0.4 | J1625-2527 and J1924-2914 |
| | 2–8 | 2022-04-23 4:53 | C43-3 | 15.1–500.2 | 46 | 75.7 | 0.5 | J1517-2422 and J1553-2422 |
| | 2–8 | 2022-04-23 7:37 | C43-3 | 15.1–500.2 | 46 | 66.5 | 0.7 | J1517-2422 and J1553-2422 |
| | 2–8 | 2022-04-23 8:57 | C43-3 | 15.1–500.2 | 46 | 47.7 | 0.8 | J1625-2527 and J1924-2914 |
| | 2–8 | 2022-04-24 6:55 | C43-3 | 15.1–500.2 | 47 | 75.2 | 0.4 | J1517-2422 and J1553-2422 |
| | 2–8 | 2022-04-24 8:15 | C43-3 | 15.1–500.2 | 47 | 56.9 | 0.4 | J1517-2422 and J1553-2422 |
| | 2–8 | 2022-04-25 8:23 | C43-3 | 15.1–500.2 | 48 | 53.8 | 0.4 | J1625-2527 and J1924-2914 |
| | 2–8 | 2022-04-26 6:54 | C43-3 | 15.1–500.2 | 43 | 73.7 | 0.6 | J1517-2422 and J1553-2422 |
| | 2–8 | 2022-04-26 8:16 | C43-3 | 15.1–500.2 | 43 | 55.0 | 0.5 | J1517-2422 and J1625-2527 |
| | 2–8 | 2022-04-27 7:06 | C43-3 | 15.1–500.2 | 45 | 70.0 | 1.0 | J1517-2422 and J1553-2422 |

**Note.** Atmospheric water is measured as precipitable water vapour (PWV); a typical value at the ALMA site is 1 mm.

---

[26] http://www.astropy.org

Table 8
SB and LB Execution UIDs for the Entire Upper Sco Sample

| For B6 USco Targets 1, 5, 6, 7, 8 | |
| --- | --- |
| SB UIDs | LB UIDs |
| A002_Xf6d51d_X4ac8 | A002_Xfb8480_Xb0ff |
| A002_Xf73ead_X172e | A002_Xfafcc0_Xf640 |
|  | A002_Xfafcc0_X9cb8 |
|  | A002_Xfafcc0_X8f72 |
|  | A002_Xfafcc0_X7f73 |
| For B6 USco Targets 2, 3, 4, 9, 10 | |
| SB UIDs | LB UIDs |
| A002_Xf6bc7b_X232f | A002_Xfaf3de_X811 |
| A002_Xf6c53e_X2699 | A002_Xfaeec7_X8a6 |
|  | A002_Xfaf3de_Xed5 |
|  | A002_Xfafcc0_X5ab |

## Appendix B
## Alternative Names

Table 9 presents other names given in the literature for the AGE-PRO sample in Upper Sco.

Table 9
Alternative Names for the AGE-PRO Upper Sco Sample

| AGE-PRO Name | 2MASS | Other Names |
| --- | --- | --- |
| USco 1 | J16120668-3010270 | TIC 95282711 (1) |
| USco 2 | J16054540-2023088 | TIC 48668314 (1) |
| USco 3 | J16020757-2257467 | EPIC 204278916 (2) |
| USco 4 | J16111742-1918285 | TIC 49166667 (1) |
| USco 5 | J16145026-2332397 | V* BV Sco (3), TIC 49440175 (1), EPIC 204130613 (2) |
| USco 6 | J16163345-2521505 | TIC 98093863 (1), EPIC 203664569 (2) |
| USco 7 | J16202863-2442087 | TIC 420918757 (1), EPIC 203848625 (2) |
| USco 8 | J16221532-2511349 | TIC 392043665 (1) |
| USco 9 | J16082324-1930009 | TIC 48871317 (1), EPIC 205080616 (2), [T64] 3 (4) |
| USco 10 | J16090075-1908526 | TIC 48913936 (1), EPIC 205151387 (2) |

**Note. References:** (1) K. G. Stassun et al. (2019), (2) S. Scaringi et al. (2016), (3) B. V. Kukarkin et al. (1971), and (4) P. S. The (1964).

# Appendix C
## Comparison between AGE-PRO Measurements for Our Upper Sco Sample and the Literature

We compared our Band 7 continuum observations and $N_2H^+$ (3–2) detections with values reported in S. A. Barenfeld et al. (2016) and D. E. Anderson et al. (2019). Half of our Upper Sco sample has published continuum observations in Band 7, but at a different frequency (340 GHz). Our continuum emission, scaled by the observed spectral index reported in Table 6, agrees with previous results. For the case of $N_2H^+$ (3–2), USco 9

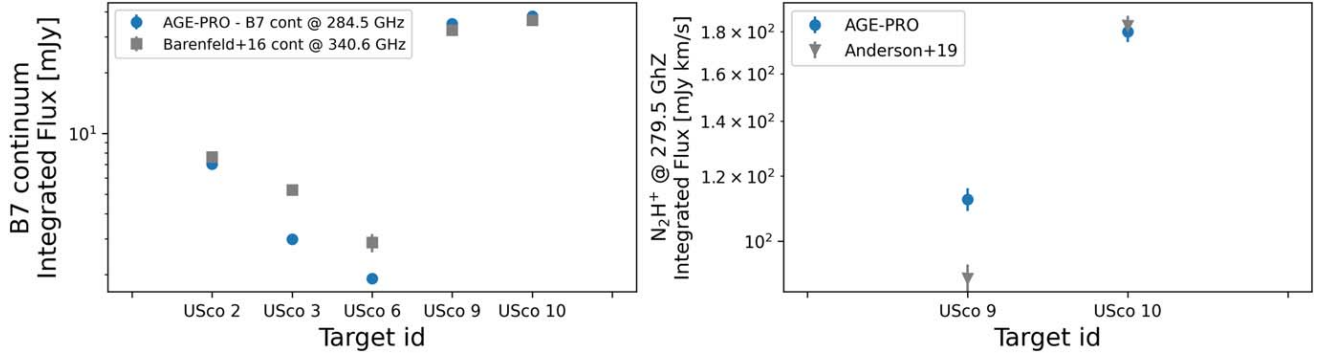

**Figure C1.** Continuum and $N_2H^+$ fluxes constrained in our AGE-PRO study at 285 GHz compared with literature values at 340 GHz (continuum) and 279.5 GHz ($N_2H^+$ 3–2). AGE-PRO results are shown in blue, and literature measurements are presented in gray (S. A. Barenfeld et al. 2016; D. E. Anderson et al. 2019).

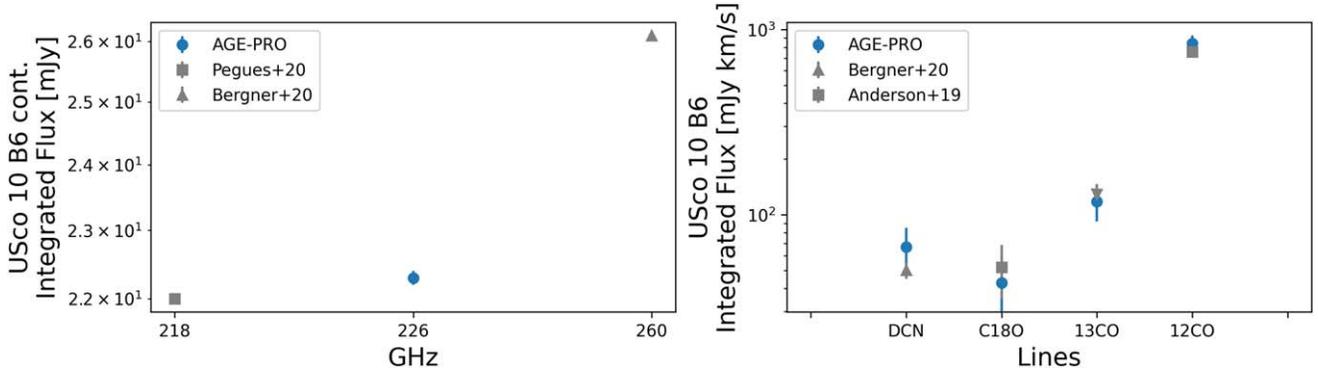

**Figure C2.** Continuum and molecular line fluxes for USco 10 compared with literature values. AGE-PRO results are shown in blue, and literature measurements are presented in gray (D. E. Anderson et al. 2019; J. B. Bergner et al. 2020; J. Pegues et al. 2020).

**Table 10**
The Flux Measurements of AGE-PRO for Continuum and $N_2H^+$ Compared with Literature Values

| Center Freq. | Line | Beam (arcsec) | Flux (mJy)(mJy km s$^{-1}$) | ID | Reference |
|---|---|---|---|---|---|
| 340.6 | cont. | ~0.35 | 7.64 ± 0.15 | USco 2 | S. A. Barenfeld et al. (2016) |
| 284.5 | cont. | 0.29 | 7.03 ± 0.03 | USco 2 | This work |
| 340.6 | cont. | ~0.35 | 5.26 ± 0.27 | USco 3 | S. A. Barenfeld et al. (2016) |
| 284.5 | cont. | 0.30 | 2.99 ± 0.04 | USco 3 | This work |
| 340.6 | cont. | ~0.35 | 2.88 ± 0.30 | USco 6 | S. A. Barenfeld et al. (2016) |
| 284.5 | cont. | 0.25 | 1.91 ± 0.03 | USco 6 | This work |
| 340.6 | cont. | ~0.35 | 32.4 ± 0.40 | USco 9 | S. A. Barenfeld et al. (2016) |
| 284.5 | cont. | 0.54 | 34.86 ± 0.12 | USco 9 | This work |
| 279.5 | $N_2H^+$ (3–2) | 0.69 | 90.0 ± 7.0 | USco 9 | D. E. Anderson et al. (2019) |
| 279.5 | $N_2H^+$ (3–2) | 0.77 | 112.4 ± 3.6 | USco 9 | This work |
| 340.6 | cont. | ~0.35 | 36.4 ± 0.30 | USco 10 | S. A. Barenfeld et al. (2016) |
| 284.5 | cont. | 0.54 | 38.04 ± 0.13 | USco 10 | This work |
| 279.5 | $N_2H^+$ (3–2) | 0.69 | 183.0 ± 8.0 | USco 10 | D. E. Anderson et al. (2019) |
| 279.5 | $N_2H^+$ (3–2) | 0.78 | 180.1 ± 5.2 | USco 10 | This work |

**Note.** The reported beam size is the geometric mean.

Table 11
Literature Flux Measurements for USco 10 in Continuum and Molecular Tracers

| Center Freq. | Line | AGE-PRO Flux (mJy km s$^{-1}$) | Beam (arcsec) | Flux (mJy)(mJy km s$^{-1}$) | References |
| --- | --- | --- | --- | --- | --- |
| 218 | cont. | ⋯ | 0.52 | 22 ± 2.2 | J. Pegues et al. (2020) |
| 260 | cont. | ⋯ | 0.43 | 26.1 ± 2.6 | J. B. Bergner et al. (2020) |
| 230 | $^{12}$CO ($J = 2$–1) | 760 ± 17.5 | 0.63 | 860 ± 21 | D. E. Anderson et al. (2019) |
| 220 | $^{13}$CO ($J = 2$–1) | 128.8 ± 18.1 | 0.62 | 118 ± 26 | J. B. Bergner et al. (2020) |
| 219 | C$^{18}$O ($J = 2$–1) | 52.0 ± 16.8 | 0.63 | 43 ± 17 | D. E. Anderson et al. (2019) |
| 217 | DCN ($J = 3$–2) | 50.3 ± 5.1 | 0.65 | 67 ± 18 | J. B. Bergner et al. (2020) |

**Note.** Uncertainties associated with continuum fluxes are determined by applying a 10% absolute flux accuracy assumption to Band 6. The reported beam size is the geometric mean.

and 10 have reported integrated fluxes, which agree within the uncertainties. Published Band 6 observations only exist for USco 10 (D. E. Anderson et al. 2019; J. B. Bergner et al. 2020; J. Pegues et al. 2020) in continuum emission, CO isotopologues, and other molecular tracers at millimeter wavelengths, as part of the ALMA project 2015.1.00964.S (PI: K. Öberg).

Figures C1 and C2 show the comparison between our AGE-PRO constraints and those reported in the literature. Tables 10 and 11 present the literature values and their observational details for comparison. Most of the identified continuum and line fluxes exhibit a concordance with the findings from previous studies. The 226 GHz continuum was not reported in previous works; therefore, our results are compared with the closest continuum Band 6 fluxes in the literature. For the integrated molecular line fluxes, all of our values are within the reported uncertainty in the literature. This confirms that our ALMA Bands 6 and 7 continuum and molecular line observations of the Upper Sco sample are in general agreement with previous literature values.

## Appendix D
## Disk Radius Comparison

Figure D1 compares our dust disk radius measured in the image plane (blue points) with the radius obtained from visibility fitting reported in the M. Vioque et al. (2025; empty squares) and the radius measured from beam-deconvolved image fitting from L. Trapman et al. (2025a; blue hexagons). In the Band 6 continuum, our observation clearly resolves USco 1, and therefore the derived $R_{90}$ and $R_{68}$ for the Band 6 continuum is close to that from the visibility fitting, and also close to the beam-deconvolved image fitting. However, all other disks are not well resolved, and therefore the radii measured in the image plane are strongly affected by the beam smearing, and hence are overestimated. The dust disk radii from the visibility fitting of the AGE-PRO disks, USco 3, 6, 9, and 10, are in agreement with the $R_{68}$ measured by N. Hendler et al. (2020).

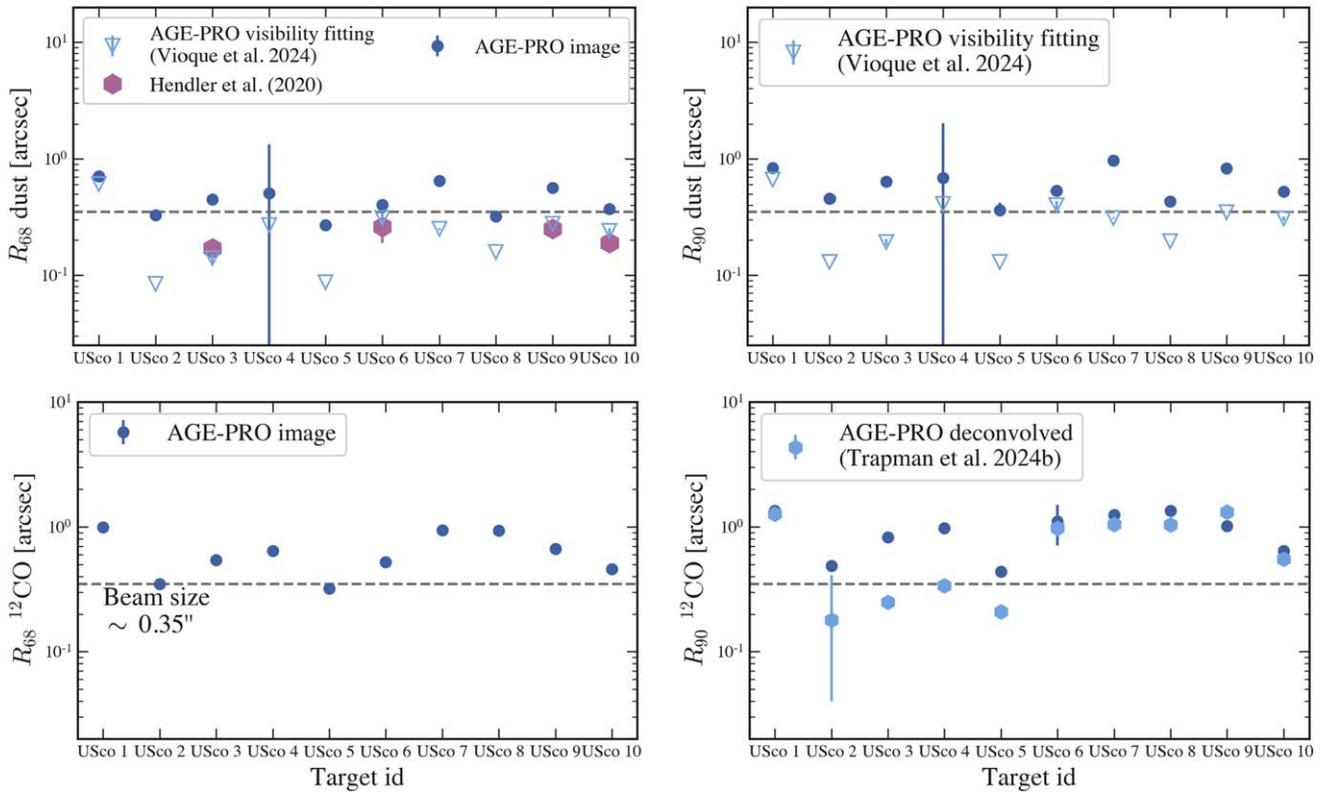

**Figure D1.** Comparison between dust and gas disk radii obtained in this work with that of AGE-PRO visibility fitting for the dust (M. Vioque et al. 2025) and beam-deconvolved image fitting for the gas (L. Trapman et al. 2025a). The $R_{68\%}$ and $R_{90\%}$ radii are on the top panels, while the bottom panels are for $R_{68\%}$ and $R_{90\%}$ gas radii measured from $^{12}$CO. Blue points correspond to the AGE-PRO disk radius measured from the images, which are convolved with the beam. Empty squares are the AGE-PRO dust disk radius from visibility fitting (M. Vioque et al. 2025) and beam-deconvolved image fitting (L. Trapman et al. 2025a). Purple hexagons are N. Hendler et al.'s (2020) dust disk radii obtained from fitting radial profile models to visibility data.

## Appendix E
## Band 6 Moment Zero Maps

Figure E1 summarizes the nondetections, marginal detections ($3\sigma <$ flux $< \sigma$), and significant detections (flux $>5\sigma$) for additional molecular lines observed in the Band 6 setup of the Upper Sco disk sample. USco 1, 3, 8, and 9 have $5\sigma$ detections in H$_2$CO ($3_{03}$–$2_{02}$), while USco 9 and 10 have $3\sigma$ and $5\sigma$ detections in DCN (3–2), respectively.

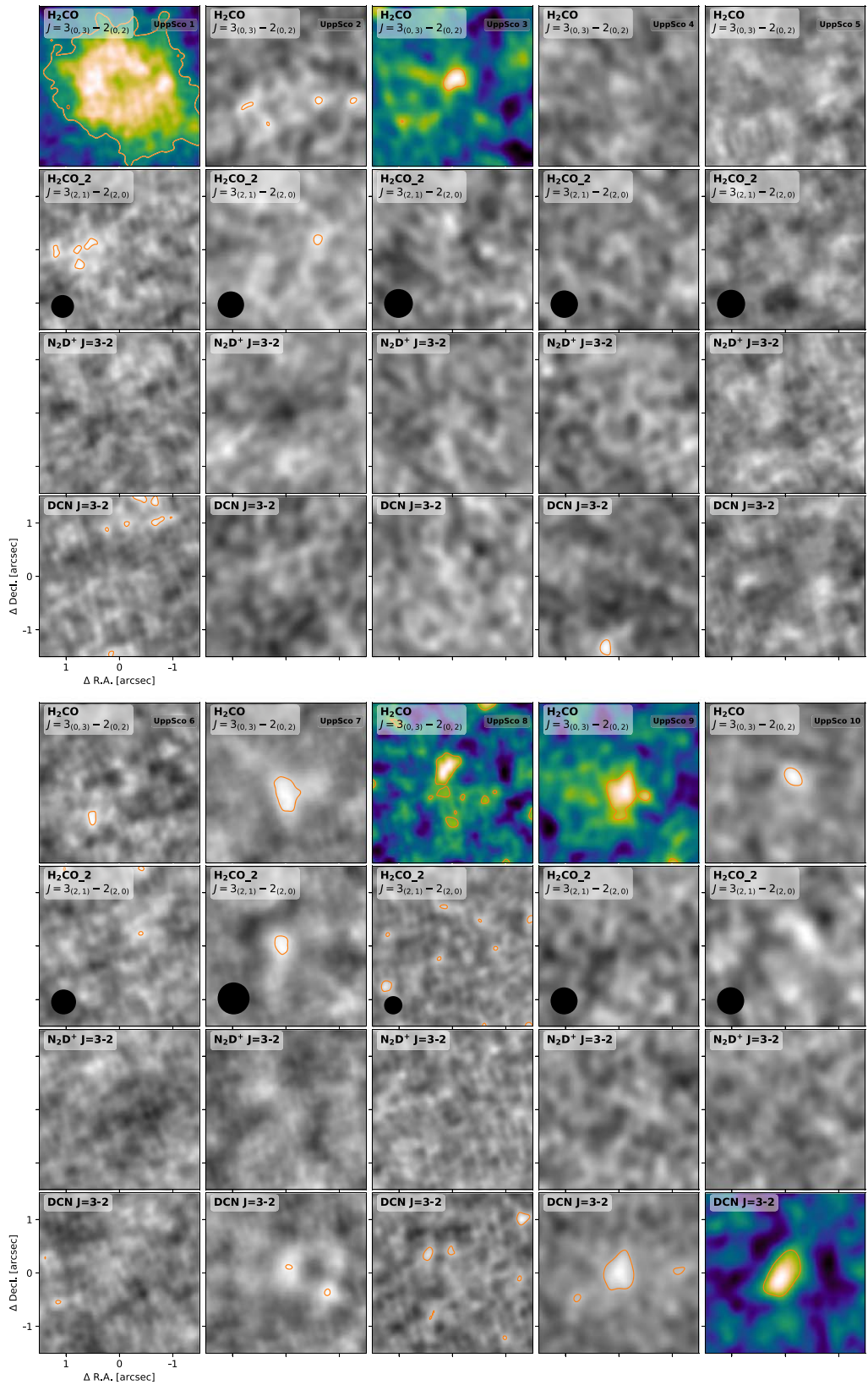

**Figure E1.** Other Band 6 lines for the Upper Sco sample. All the images use the robust 1.0 with a circularized beam. Contour overlaps in rainforest maps show $5\sigma$ detections, while contours over gray maps show $3\sigma$ or nondetections. The continuum color scale is normalized to peak intensity.

## Appendix F
## CO Channel Maps

The $^{12}$CO($J = 2$–1) emission is detected on all Upper Sco disks. Figures F1–F5 show the channel maps of the CO line emission from $V_{\rm LSRK} = -1.0$ to $+7.0$ km s$^{-1}$.

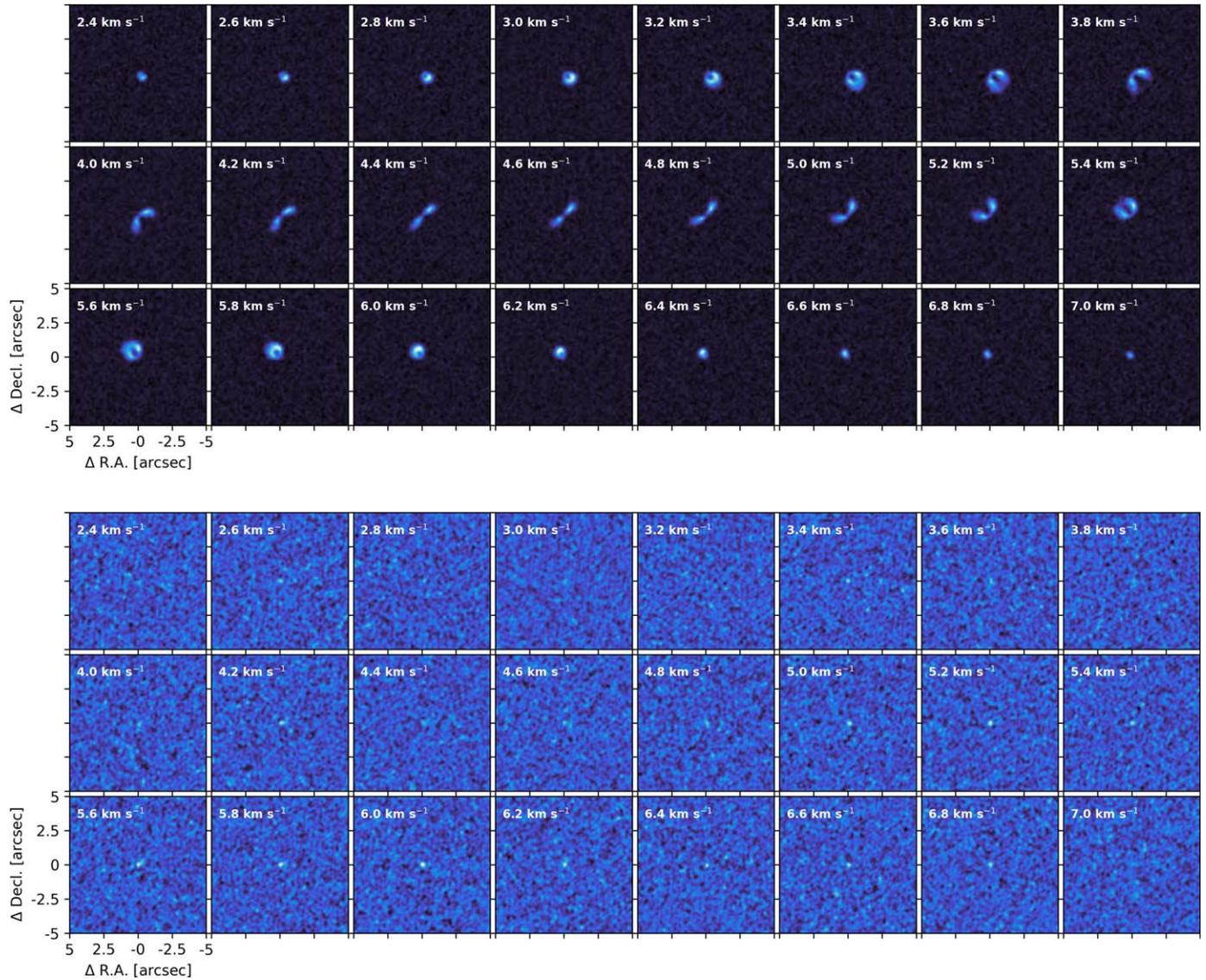

**Figure F1.** Channel maps of the $^{12}$CO ($J = 2$–1) emission in the USco 1 (top) and USco 2 (bottom) disks.

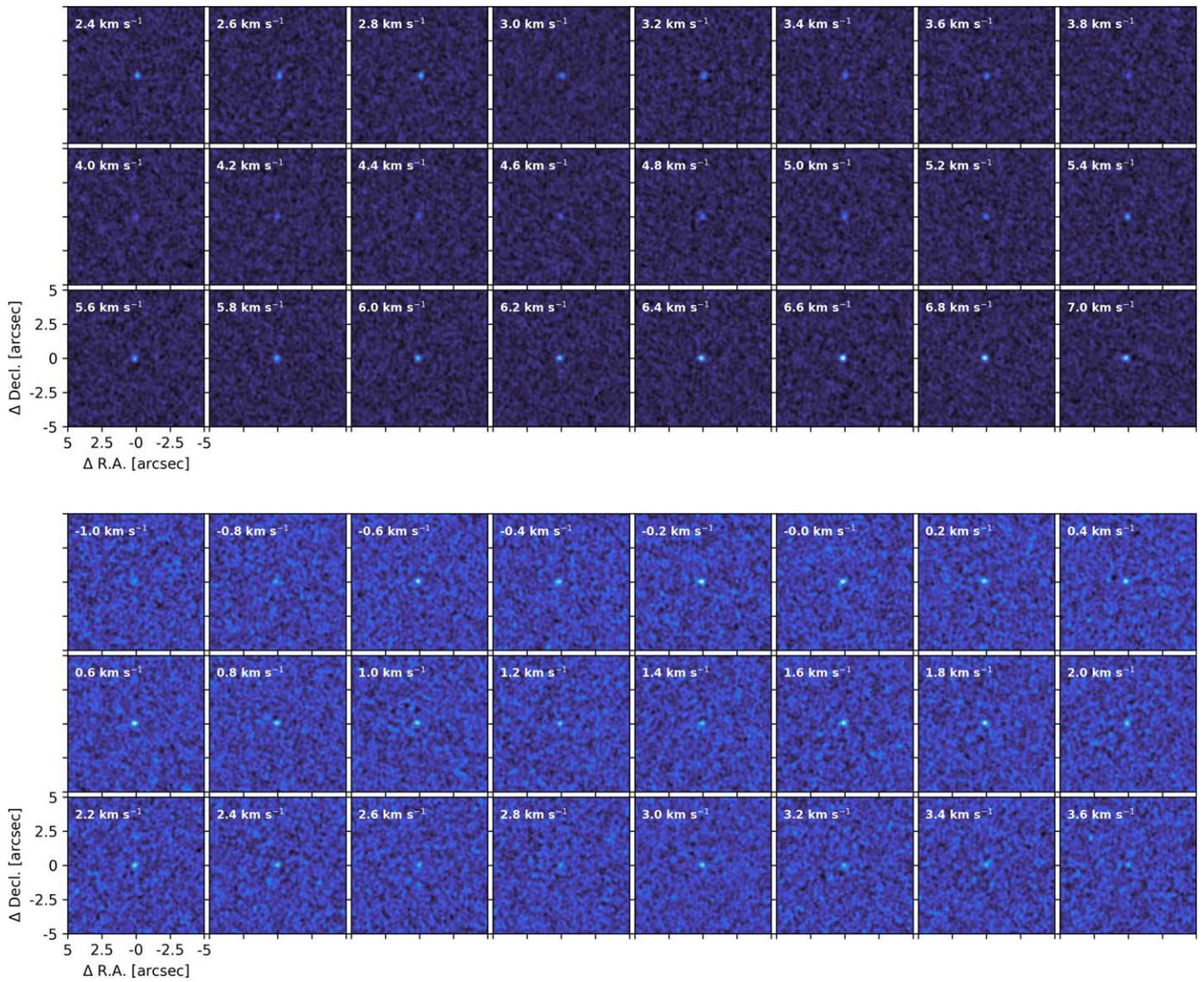

**Figure F2.** Channel maps of the $^{12}$CO ($J = 2$–1) emission in the USco 3 (top) and USco 4 (bottom) disks.

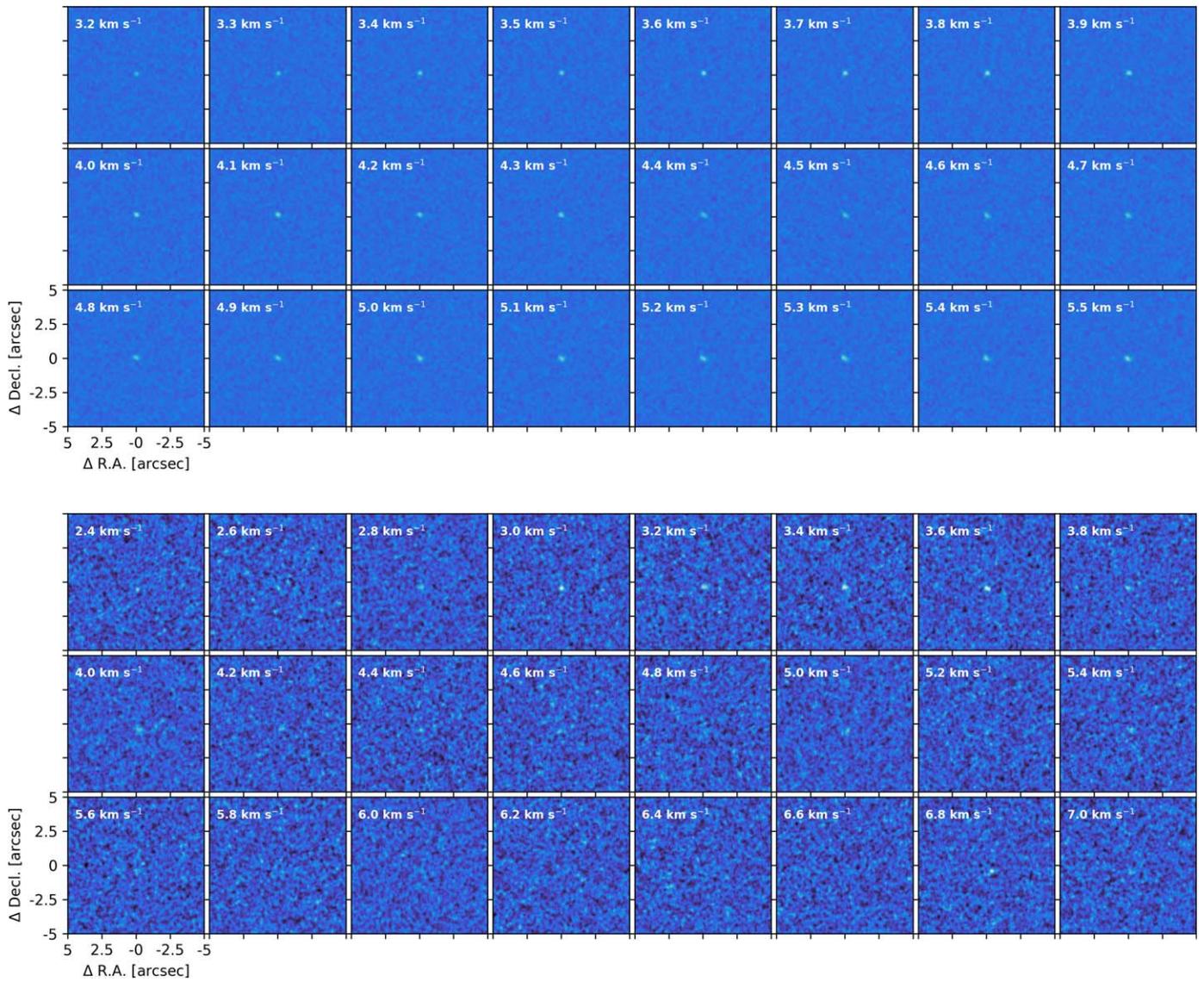

**Figure F3.** Channel maps of the CO ($J = 2$–1) emission in the USco 5 (top) and 6 (bottom) disks.

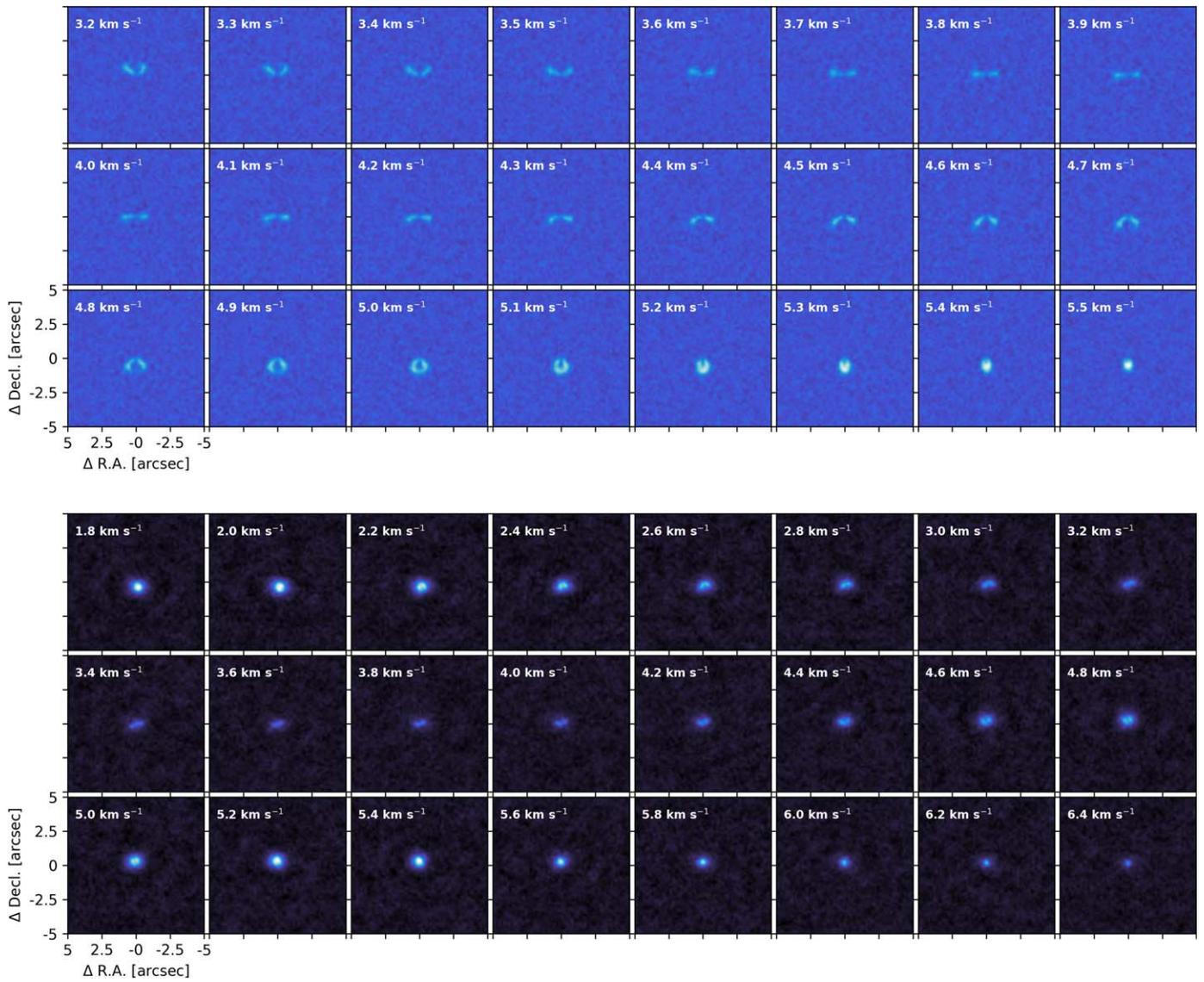

**Figure F4.** Channel maps of the CO ($J = 2$–$1$) emission in the USco 7 (top) and 8 (bottom) disks.

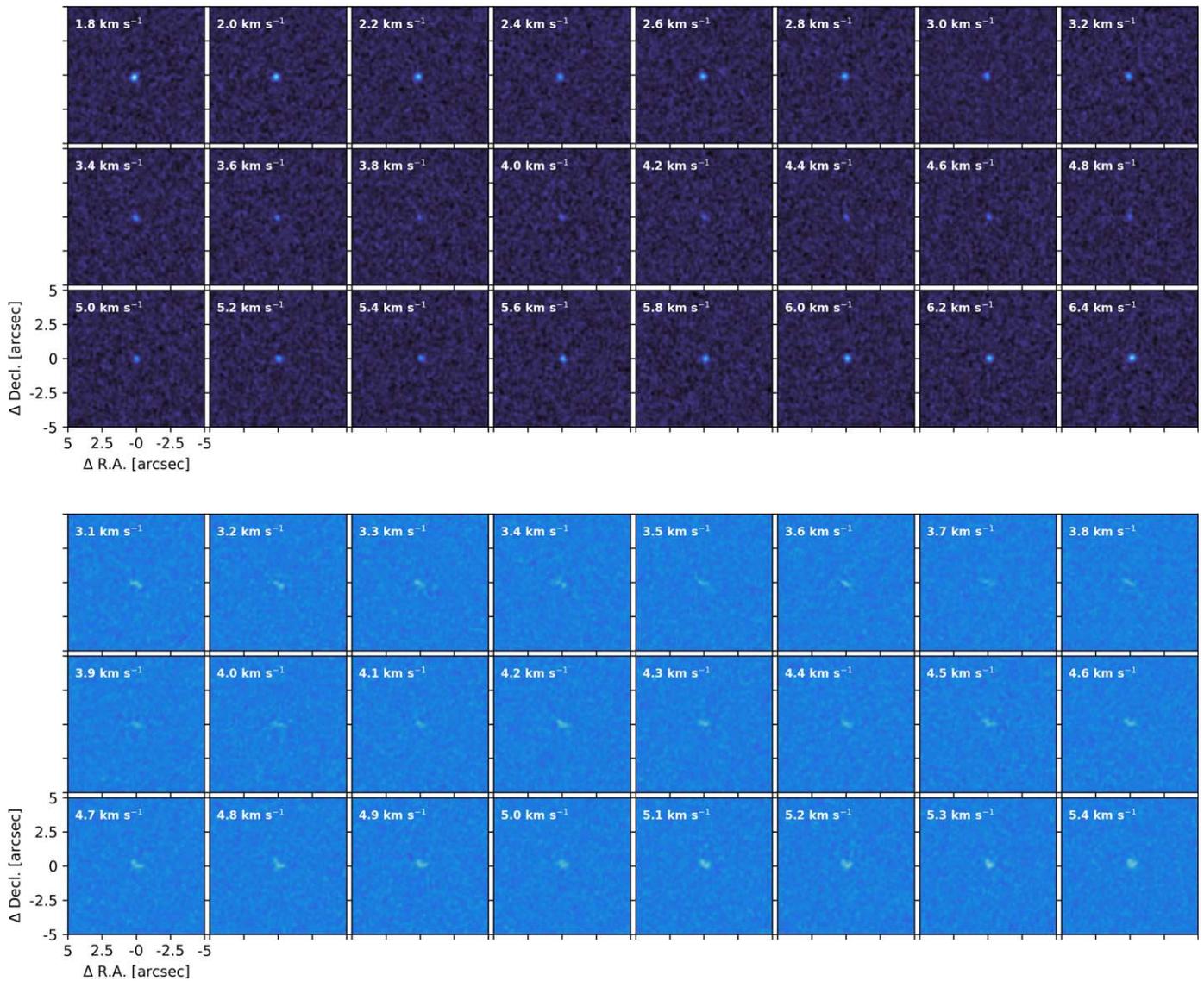

**Figure F5.** Channel maps of the CO ($J = 2$–1) emission in the USco 9 (top) and 10 (bottom) disks.

### ORCID iDs

Carolina Agurto-Gangas ⓘ https://orcid.org/0000-0002-7238-2306
Laura M. Pérez ⓘ https://orcid.org/0000-0002-1199-9564
Anibal Sierra ⓘ https://orcid.org/0000-0002-5991-8073
James Miley ⓘ https://orcid.org/0000-0002-1575-680X
Ke Zhang ⓘ https://orcid.org/0000-0002-0661-7517
Ilaria Pascucci ⓘ https://orcid.org/0000-0001-7962-1683
Paola Pinilla ⓘ https://orcid.org/0000-0001-8764-1780
Dingshan Deng ⓘ https://orcid.org/0000-0003-0777-7392
John Carpenter ⓘ https://orcid.org/0000-0003-2251-0602
Leon Trapman ⓘ https://orcid.org/0000-0002-8623-9703
Miguel Vioque ⓘ https://orcid.org/0000-0002-4147-3846
Giovanni P. Rosotti ⓘ https://orcid.org/0000-0003-4853-5736
Nicolas Kurtovic ⓘ https://orcid.org/0000-0002-2358-4796
Lucas A. Cieza ⓘ https://orcid.org/0000-0002-2828-1153
Rossella Anania ⓘ https://orcid.org/0009-0004-8091-5055
Benoît Tabone ⓘ https://orcid.org/0000-0002-1103-3225
Kamber Schwarz ⓘ https://orcid.org/0000-0002-6429-9457
Michiel R. Hogerheijde ⓘ https://orcid.org/0000-0001-5217-537X
Estephani E. TorresVillanueva ⓘ https://orcid.org/0000-0001-9961-8203
Dary A. Ruiz-Rodriguez ⓘ https://orcid.org/0000-0003-3573-8163
Camilo González-Ruilova ⓘ https://orcid.org/0000-0003-4907-189X